\documentclass[journal, final]{IEEEtran}

\usepackage{graphicx}
\usepackage{amssymb}
\usepackage{amsmath}
\usepackage{multirow}
\usepackage{amstext}
\usepackage{bm}
\usepackage{subfigure,color}
\usepackage{cite}
\usepackage{ulem}
\usepackage{cancel}
\newtheorem{theorem}{Theorem}[section]
\newtheorem{definition}{Definition}[section]
\newtheorem{lemma}{Lemma}[section]
\newtheorem{corollary}{Corollary}[section]
\newtheorem{remark}{Remark}[section]
\newtheorem{proposition}{Proposition}[section]
\newtheorem{conjecture}{Conjecture}[section]

\begin{document}
\title{Matrix Infinitely Divisible Series: Tail Inequalities and Their Applications}

\author{Chao~Zhang~\IEEEmembership{Member,~IEEE,} Xianjie~Gao, Min-Hsiu Hsieh$^*$~\IEEEmembership{Senior Member,~IEEE,} Hanyuan~Hang, Dacheng Tao,~\IEEEmembership{Fellow,~IEEE}
\thanks{C.~Zhang and X.~Gao are with the School of Mathematical Sciences, Dalian University of Technology, Dalian, Liaoning, 116024, P.R. China. E-mail: chao.zhang@dlut.edu.cn, xianjiegao@foxmail.com.}
\thanks{M.-H.~Hsieh is with Centre for Quantum Software and Information, University of Technology Sydney, Sydney NSW 2007, Australia. E-mail: Min-Hsiu.Hsieh@uts.edu.au.}
\thanks{H.~Hang is with the AI Lab, Samsung Research China - Beijing, Beijing 100028, P. R. China E-mail: hanyuan.hang@samsung.com.}
\thanks{D. Tao is with the UBTECH Sydney Artificial Intelligence Centre and the School of Computer Science, in the Faculty of Engineering, at The University of Sydney, 6 Cleveland St, Darlington, NSW 2008, Australia. E-mail: dacheng.tao@sydney.edu.au.}
\thanks{CZ is supported by the National Natural Science Foundation of China: 61473328 and 11401076; and the Fundamental Research Funds for the Central Universities: DUT17LK46. MH is supported by an ARC Future Fellowship under Grant FT140100574. DT is supported by Australian Research Council Project FL-170100117.}
\thanks{$^*$Corresponding author}
}


\maketitle

\begin{abstract}

In this paper, we study tail inequalities of the largest eigenvalue of a matrix infinitely divisible (i.d.) series, which is a finite sum of fixed matrices weighted by i.d. random variables. We obtain several types of tail inequalities, including Bennett-type and Bernstein-type inequalities. This allows us to further bound the expectation of the spectral norm of a matrix i.d. series. Moreover, by developing a new lower-bound function for $Q(s)=(s+1)\log(s+1)-s$ that appears in the Bennett-type inequality, we derive a tighter tail inequality of the largest eigenvalue of the matrix i.d. series than the Bernstein-type inequality when the matrix dimension is high. The resulting lower-bound function is of independent interest and can improve any Bennett-type concentration inequality that involves the function $Q(s)$. The class of i.d. probability distributions is large and includes Gaussian and Poisson distributions, among many others. Therefore, our results encompass the existing work \cite{tropp2012user} on matrix Gaussian series as a special case. Lastly, we show that the tail inequalities of a matrix i.d. series have applications in several optimization problems including the chance constrained optimization problem and the quadratic optimization problem with orthogonality constraints. {In addition, we also use the resulting tail bounds to show that random matrices constructed from i.d. random variables satisfy the restricted isometry property (RIP)  when it acts as a measurement matrix in compressed sensing.}

\end{abstract}

\begin{IEEEkeywords}
Random matrix, tail inequality, infinitely divisible distribution, largest eigenvalue, optimization, {restricted isometry property}, compressed sensing
\end{IEEEkeywords}

\section{Introduction}\label{sec:intro}

Random matrices have been widely used in many machine learning and information theory problems, {\it e.g.,} compressed sensing \cite{Andersson2014On,Dirksen2016On,Vehkaper2016Analysis}, coding theory \cite{Wei2015From}, kernel method \cite{Jin2013Improved}, estimation of covariance matrices \cite{Yazdian2016Eigenvalue,Couillet2012Robust}, and quantum information theory \cite{Cheng20150563, doi:10.1063/1.5000846, Cheng:2016:LUQ:3179466.3179470, ahlswede2002strong}. In particular, sums of random matrices and the tail behavior of their extreme eigenvalues (or singular values) are of significant interest in theoretical studies and practical applications ({\it cf.}~\cite{tropp2015introduction}). Ahlswede and Winter  presented a large-deviation inequality for the extreme eigenvalues of sums of random matrices \cite{ahlswede2002strong}.
Tropp improved upon their results using Lieb's concavity theorem \cite{tropp2012user}. Hsu {\it et al.} provided tail inequalities for sums of random matrices that depend on intrinsic dimensions instead of explicit matrix dimensions \cite{hsu2012tail}. By introducing the concept of effective rank, Minsker extended Bernstein's concentration inequality for random matrices \cite{minsker2017on}  and refined the results in \cite{hsu2012tail}. There have also been many other works on the eigenproblems of random matrices ({\it cf.} \cite{meckes2004concentration,mackey2014matrix,paulin2013deriving,vershynin2010introduction,Chiani2017On}), and the list provided here is  incomplete.


A simple form of sums of random matrices can be expressed as $\sum_k\xi_k{\bf A}_k$ with random variables $\xi_k$ and fixed matrices ${\bf A}_k$. This form has played an important role in recent works on neural networks \cite{zhao17theoretical}, kernel methods \cite{choromanski2016recycling} and deep learning \cite{cheng2015exploration}, where the original weighted  (or projection) matrices can be replaced with structured random matrices, such as circulant and Toeplitz matrices with Gaussian or Bernoulli entries. Note that these two distributions, along with uniform distributions and Rademacher distributions, belong to the family of sub-Gaussian distributions\footnote{A random variable $\xi$ is said to be sub-Gaussian if its moment generating function (mgf) satisfies $\mathbb{E}[{\rm e}^{\theta \xi}]\leq{\rm e}^{\theta^2c^2}$ ($\theta\in \mathbb{R}$), where $c$ is an absolute constant.}, and many techniques dedicated to sub-Gaussian random matrices have been developed ({\it e.g.,} \cite{tropp2012user,hsu2012tail}). However, to the best of our knowledge, random matrix research beyond that is still very limited.

The tail behavior of $\|\sum_k\xi_k{\bf A}_k\|$, where $\|{\bf A}\|$ stands for the spectral norm of the matrix ${\bf A}$, is strongly related to several optimization problems, including the Procrustes problem and the quadratic assignment problem ({\it cf.} \cite{nemirovski2007sums,so2011moment}). 
Nemirovski analyzed efficiently computable solutions to these optimization problems \cite{nemirovski2007sums}, and showed that the tail behavior of $\|\sum_k\xi_k{\bf A}_k\|$ provides answers to 1) the safe tractable approximation of chance constrained linear matrix inequalities, and 2) the quality of semidefinite relaxations of a general quadratic optimization problem. He also proved a tail bound for $\|\sum_k\xi_k{\bf A}_k \|$, where $\{\xi_k\}$ obey either distributions supported on $[-1,1]$ or Gaussian distributions with {\it unit} variance, and presented a conjecture for the  ``optimal" expression of the tail bound \cite{nemirovski2007sums}. Anthony So applied the non-commutative Khintchine's inequality to achieve a solution to Nemirovski's conjecture \cite{so2011moment}. Note that the aforementioned results assume that $\{\xi_k\}$ obey distributions supported on $[-1,1]$ or Gaussian distributions with {\it unit} variance. These assumptions will not always be satisfied in practice, and it is advantageous to explore whether these efficiently computable optimization solutions would also hold in a broader setting. We answer this question in the affirmative in this paper.

In this work, we study and prove tail bounds for the random matrix $\sum_k\xi_k{\bf A}_k$, where random variables $\{\xi_k\}$ are infinite divisible distributions. The class of infinitely divisible (i.d.) distributions includes Gaussian distributions, Poisson distributions, stable distributions and compound Poisson distributions as special cases ({\it cf.} \cite{bose2002contemporary,kyprianou2006introductory}). In recent years, techniques developed for i.d.~distributions have been employed in important applications in the fields of image processing \cite{chainais2005multi} and kernel methods \cite{nishiyama2016characteristic}. Note that  there is no intersection between sub-Gaussian distributions and i.d. distributions except for Gaussian distributions ({\it cf.} Lemma 5.5 of \cite{vershynin2010introduction}). We therefore believe that our works on random matrix with respect to i.d. distributions will complement earlier results for sub-Gaussian distributions and provide useful applications in the fields of learning and optimization, and beyond.


\subsection{Overview of the Main Results}

There are three main contributions of this paper: 1) we obtain tail inequalities for the largest eigenvalue of the matrix infinitely divisible (i.d.) series $\sum_k\xi_k{\bf A}_k$, where the $\xi_k$ are i.d. random variables; 2) we construct a piecewise function to bound the function $Q(s)=(s+1)\log(s+1)-s$ from below when $s\in(0,c]$ for any given $1<c<+\infty$, and the new lower bound function is the  tightest up to date; and 3) we show that the tail inequalities of matrix i.d. series provide efficiently computable solutions to several optimization problems.

First, we develop a matrix moment-generating function (mgf) bound for i.d. distributions as the starting point for deriving the subsequent tail inequalities for the matrix i.d. series. Then, we derive the tail inequality given in \eqref{eq:tail1} for the matrix i.d. series, which is difficult to compute because of the integral of an inverse function. Therefore, by introducing the additional condition that the L\'evy measure has a bounded support, we simplify the aforementioned result into a Bennett-type tail inequality [{\it cf.} \eqref{eq:tail2}] that contains the function $Q(s)=(s+1)\log(s+1)-s$, and we also replace $Q(s)$ with $B(s)=\frac{s^2}{2(1+s/3)}$ to obtain a Bernstein-type tail inequality [{\it cf.} \eqref{eq:tail3}] for the matrix i.d. series. In addition, we bound the expectation of the spectral norm of the matrix i.d. series.

Since $B(s)$ cannot bound $Q(s)$ from below sufficiently tightly when $s$ is large ({\it cf.} Fig. \ref{fig:QBT}), we introduce another function $H_P(s)$ [{\it cf.} \eqref{eq:new1}] to bound $Q(s)$ from below more tightly than $B(s)$ when $s\in(0.8831,c]$ for any $1<c<+\infty$ ({\it cf.} Remark \ref{rem:8831}). Although $H_P(s)$ is a piecewise function, all sub-functions of $H_P(s)$ share the simple form $\beta_0s^{\tau_n}$ (where $\beta_0=2\log2-1$) and thus have a low computational cost, and the subdomains of $H_P(s)$ can be arbitrarily selected as long as points $1$ and $c$ are included in the ordered sequence $P$ as the smallest and largest elements, respectively. Based on $H_P(s)$ (especially with $P=\{1,c\}$), we obtain another type of tail inequality for matrix i.d. series that is tighter than the Bernstein-type result given in \eqref{eq:tail3} when $\frac{Rt}{\rho(\sigma^2+V)}>0.8831$.\footnote{In general, the tail inequality $\mathbb{P}\{\xi>t\}$ describes the probability characteristics of the event in which the value of a random variable $\xi$ is greater than a given positive constant $t$. Consequently, the tail inequality provides more useful information in the case of $\frac{Rt}{\rho(\sigma^2+V)}>0.8831$ than in the case of $\frac{Rt}{\rho(\sigma^2+V)}\leq 0.8831$.} We show that the tail result based on $H_P(s)$ provides a tighter upper bound on the largest eigenvalue of a matrix i.d. series than is possible with the Bernstein-type result when the matrix dimension is high. The results regarding $Q(s)$ and $H_P(s)$ are applicable for any Bennett-type concentration inequality that involves the function $Q(s)$.

Using the resulting tail bounds for random i.d. series, we study the properties of two optimization problems: chance constrained optimization problems and quadratic optimization problems with orthogonality constraints, which covers several well-studied optimization problems as special cases, {\it e.g.,} the Procrustes problem and the quadratic assignment problem. Although these problems have been exhaustively explored in the works \cite{nemirovski2007sums,so2011moment}, their results are built under the assumption that $\xi_k$ obey either distributions  supported on $[-1,1]$ or Gaussian distributions with {\it unit} variance, which restricts the feasibility of the results in practical problems that do not satisfy the assumption. By using the tail inequalities for random i.d. series to resolve an extension of Nemirovski's conjecture ({\it cf.} Conjecture \ref{conjecture}), we show that the results obtained in \cite{nemirovski2007sums,so2011moment} are also valid in the i.d. scenario, where $\xi_k$ obey i.d. distributions instead of distributions supported on $[-1,1]$ or Gaussian distributions. 

{As an application of the resulting tail bounds in compressed sensing, we explore the restricted isometry property (RIP) of a random i.d. series \cite{candes2006near,mendelson2008uniform}. In particular, we show that if a measurement matrix ${\bf A}$ can be expressed as a random i.d. series, {\it i.e.,} ${\bf A} = \sum_{k=1}^K \xi_k{\bf A}_k$, where $\xi_k$ $(1\leq k\leq K)$ obey a generalized gamma convolution distribution, a subclass of i.d. distributions, it satisfies the RIP with a high probability. Our result hence extends earlier results that the random circulant and Toeplitz matrices, constructed from the random Gaussian (or Bernoull) series, serve as good measurement matrices in compressed sensing \cite{rauhut2009circulant,haupt2010toeplitz}.}

The remainder of this paper is organized as follows. Section \ref{sec:definition} introduces necessary preliminaries on i.d. distributions and Section \ref{sec:basic} presents the main results of this paper. In Section \ref{sec:conjecture}, we study the application of random i.d. series in a number of optimization problems. {In Section \ref{sec:rip}, we discuss the RIP of random i.d. series in compressed sensing.} Section \ref{sec:con} concludes the paper. In the appendix, we provide a detailed introduction to the L\'evy measure (part \ref{app:levy}) and prove the main results of this paper (part \ref{app:proof}). 


\section{Preliminaries on Infinitely Divisible Distributions}\label{sec:definition}

In this section, we first introduce several definitions related to infinitely divisible (i.d.) distributions and then present the matrix mgf inequality for i.d. distributions.

\subsection{Infinitely Divisible Distributions}

We begin with the definition of the L\'{e}vy measure.
\begin{definition}[L\'evy Measure]\label{def:levy.meausre}
A Borel measure $\nu$ is said to be a L\'{e}vy measure if it satisfies 
\begin{equation}\label{L.Meas.}
    \int_{\mathbb{R}}\min\{ u^2,1\}\nu(du)<\infty \;\;\mbox{and $\nu(\{0\})=0$.}
\end{equation}
\end{definition}
The L\'{e}vy measure describes the expected number of jumps of a certain height in a time interval of {\it unit} length, and a more detailed discussion is given in Appendix \ref{app:levy} for completeness.

A random variable $\xi\in\mathbb{R}$ has an i.d. distribution if for any $n>1$, there exists a sequence $\{\xi_n^{(1)}, \cdots, \xi^{(n)}_n\}$ of independent and identically distributed (i.i.d.) random variables such that $\xi$ has the same distribution as $\xi_n^{(1)}+\cdots+\xi_n^{(n)}$. The following theorem provides a sufficient and necessary condition for i.d. distributions:


\begin{theorem}[L\'{e}vy-Khintchine Theorem]\label{thm:cf}
A real-valued random variable $\xi$ is i.d. if and only if there exists a triplet
$(b,\sigma^2,\nu)$ such that for any $\theta\in\mathbb{R}$, the characteristic  function of $\xi$ is of the form
\begin{equation}\label{eq:levy.formula}
\mathbb{E}\{ e^{i \theta\xi}\}=\exp\Big(ib\theta -\frac{\sigma^2\theta^2}{2}+\int_\mathbb{R}\big({\rm e}^{i\theta u}-1-i\theta u\textbf{1}_{|u|< 1}\big)\nu (du)\Big),
\end{equation}
where $b\in\mathbb{R}$, $\sigma \geq 0$ and $\nu$ is a L\'{e}vy measure. 
\end{theorem}
This theorem states that an i.d. distribution can be characterized by the triplet $(b,\sigma^2,\nu)$ (see also Refs.~\cite{bose2002contemporary,kyprianou2006introductory}).

\subsection{Matrix Inequalities for Infinitely Divisible Distributions}

Let the symbol $\preceq$ denote the semidefinite order on self-adjoint matrices, i.e., $\bf A \preceq \bf B$ means that the matrix $\bf B -\bf A$ is positive semi-definite. For any real functions $f$ and $g$, the transfer rule states that if $f(a)\leq g(a)$ for any $a\in I$, then $f({\bf A})  \preceq g({\bf A}) $ when the eigenvalues of the semidefinite matrix ${\bf A}$ lie in $I$. Let $\lambda_{\max}({\bf A})$ stand for the largest eigenvalue of a self-adjoint matrix ${\bf A}$. 

Below, we present the matrix mgf bound for i.d. distributions as the starting point for deriving the desired tail results for matrix i.d. series.
\begin{lemma}\label{lem:id.mgf}
Let $\xi$ be an i.d. random variable with the triplet $(b,\sigma^2,\nu)$, and suppose that $\mathbb{E} \xi = 0$. Let $M:=\sup \{\theta\geq0:\,\mathbb{E}\{{\rm e}^{\theta |\xi|}\}<+\infty\}$. Given a fixed self-adjoint matrix ${\bf A}$ with $\lambda_{\max}({\bf A}) \leq 1$, it holds that, for any {$0<\theta < M$},
\begin{equation}\label{eq:mgf1}
\mathbb{E}\{ {\rm e}^{\xi\theta  {\bf A}} \}\preceq {\rm e}^{\Phi(\theta)\cdot {\bf A}^2},
\end{equation}
where 
\begin{equation}\label{eq:mgf2}
\Phi(\theta) := \frac{\sigma^2 \theta^2}{2}
+\int_{\mathbb{R}}\left(\mathrm{e}^{\theta|u|}-\theta |u|-1\right)\nu(du).
\end{equation}
\end{lemma}
The proof of this lemma is given in Appendix~\ref{app.lem1}.
Note that if the L\'evy measure $\nu$ is the {\it zero} measure, then \eqref{eq:mgf1} is analogous to the result: $\mathbb{E}\{{\rm e}^{\xi\theta{\bf A}}\}={\rm e}^{\theta^2{\bf A}^2/2}$, $\theta\in\mathbb{R}$, when $\xi$ is Gaussian ({\it cf.} \cite[Lemma 4.3]{tropp2012user}). In addition, the requirement that $M>0$ actually causes the above result to exclude the heavy-tailed distributions.\footnote{A random variable $\xi$ is said to obey a heavy-tailed distribution if $\mathbb{E}\{{\rm e}^{\theta \xi}\}=\infty$ holds for any $\theta > 0$.} However, as mentioned by Houdr\'e \cite{houdre2002remarks}, there are still many specific i.d. distributions that satisfy such a requirement, {\it e.g.,} Poisson distributions, geometric distributions, negative binomial distributions, Gamma distributions and compound Poisson distributions.


\section{Tail Inequalities for Matrix Infinitely Divisible Series}\label{sec:basic}

Let $\xi_1,\cdots,\xi_K$ be independent centered infinitely divisible (i.d.)~random variables with the triplet $(b,\sigma^2,\nu)$. Let ${\bf A}_1,\cdots,{\bf A}_K$ be fixed d-dimensional self-adjoint matrices with $\lambda_{\max}({\bf A}_k)\leq 1$, $k=1,\cdots,K$, throughout the rest of the paper (unless otherwise stated). The target of this section is the random matrix ${\bf{A}}=\sum_{k=1}^K \xi_k {\bf{A}_k}$ that is constructed from i.d.~random variables $\xi_1,\cdots,\xi_K$.  We first present two types of tail inequalities for matrix i.d. series: Bennett-type and Bernstein-type inequalities. By analyzing the characteristics of the function $Q(s) = (s+1)\cdot \log(s+1)-s$ that appears in the Bennett-type result, we introduce a piecewise function $H(s)$ to bound $Q(s)$ from below and thus obtain a new tail inequality for matrix i.d. series. We also study the upper bound of the expectation of $\|\sum_k \xi_k {\bf A}_k \|$.

\subsection{Tail Inequalities for Matrix Infinitely Divisible Series}

Denote $M:=\sup \{\theta\in\mathbb{R}:\,\mathbb{E}\,{\rm e}^{\theta |\xi|}<+\infty\}$. Define $\rho:= \lambda_{\max}\big(\sum_{k=1}^K  {\bf A}_k^2\big)$ {and let
\begin{equation*}
\alpha(s):=\sigma^2s+\int_\mathbb{R} |u|({\rm e}^{s |u|}-1) \nu(du),\quad 0<s<M.
\end{equation*}}
By using the matrix mgf bound \eqref{eq:mgf1}, we obtain the first tail inequality for the matrix i.d. series $\sum_{k=1}^K \xi_k {\bf A}_k$. 




\begin{theorem}\label{thm:tail} For any $0<t< \rho\alpha(M^-)$, we have
 \begin{align}\label{eq:tail1}
  &\mathbb{P}\left\{\lambda_{\max}\left(\sum_k \xi_k {\bf A}_k\right)> t\right\}\nonumber\\
  &\leq {{d}}\exp\left(-\rho\cdot\int_{0}^{t/\rho}\alpha^{-1}(s)ds\right),
\end{align}
where $\alpha(M^-):= \lim_{s\uparrow M} \alpha(s)$ and $\alpha^{-1}(s)$ is the inverse of $\alpha(s)$.
\end{theorem}
\begin{IEEEproof}
The proof of this theorem is given in Appendix~\ref{app.thm1}.
\end{IEEEproof}
\begin{remark}
 Since the matrices ${\bf A}_k$ ($1\leq k\leq K$) are self-adjoint, the matrix $\sum_k  {\bf A}_k^2$ is self-adjoint and positive semidefinite. Therefore, $\rho$ is non-negative and the above result is non-trivial.  
\end{remark}
 
\begin{remark}\label{rem:spectral}
As pointed out in \cite[Section 2.6]{tropp2012user}, 
\begin{equation*}
2\mathbb{P}\left\{\lambda_{\max}\left(\sum_k \xi_k {\bf A}_k\right)> t\right\}=\mathbb{P}\left\{\Big\|\sum_k \xi_k {\bf A}_k\Big\|> t\right\},
\end{equation*}
Theorem~\ref{thm:tail} can also be used to study the tail behavior of the spectral norm (or the largest singular value) of random i.d. series.
\end{remark}

\begin{remark}\label{rem:tvalue}
As shown in the proof of Theorem 3.1, setting $t< \rho\alpha(M^{-})$ aims to guarantee that the solution $\theta=\alpha^{-1}(t/\rho)$ to the optimization problem $\min_{0<\theta<M}\left\{\rho\cdot\Phi(\theta)-\theta\cdot t\right\}$ lies in the interval $(0,M)$. Actually, when $t\geq \rho\alpha(M^{-})$, according to the convexity of $\rho\Phi(\theta)-\theta t$ w.r.t. $\theta>0$ and the monotonicity of $\alpha^{-1}(s)$ ($s>0$), the solution to the optimization problem is $\theta = M$. Thus, for any $t\geq\rho\alpha(M^{-})$, we have
\begin{equation*}
\mathbb{P}\left\{\lambda_{\max}\left(\sum_k \xi_k {\bf A}_k\right)> t\right\}\leq d \exp\left(\rho \Phi(M)-M  t\right),
\end{equation*}
which has the same order $O({\rm e}^{-t})$ as the second sub-function of the Bernstain-type result (iii) in Theorem 6.1 of \cite{tropp2012user}.
\end{remark}


Considering the difficulties that arise in computing the function $\alpha(s)$ and its inverse $\alpha^{-1}(s)$, we introduce the additional condition that $\nu$ has a bounded support{, {\it i.e.,} there exists a positive constant $a<\infty$ such that $\nu((-\infty,-a)\cup(a,\infty)) =0$ and $\nu([-a,a]) \not =0$.} Let 
\begin{equation}\label{eq_R}
R=\inf\{a>0:\nu(\{u:|u|>a\})=0\}.
\end{equation} 
It follows that $R<\infty$. Then Theorem~\ref{thm:tail} has the following simple form under this additional assumption.
\begin{corollary}\label{cor:tail1}
If $\nu$ has a bounded support with $R$ defined in (\ref{eq_R}), then for any $t>0$,
\begin{align}\label{eq:tail2}
&\mathbb{P}\left\{\lambda_{\max}\left(\sum_k \xi_k {\bf A}_k\right)> t\right\}\nonumber\\
\leq& d\cdot \exp\left( - \frac{\rho(\sigma^2 +V)}{R^2} \cdot Q \left(\frac{Rt}{\rho(\sigma^2+V)}\right) \right),
\end{align}
where $V:= \int_{\mathbb{R}}|u|^2 \nu(du)$, and
 \begin{equation}\label{eq:Gamma}
Q(s) := (1+s)\cdot \log(1+s)-s.
\end{equation}

\end{corollary} 
The proof of this corollary is given in Appendix~\ref{app.corr1}.

Roughly speaking, the condition that $\nu$ has a bounded support means that large jumps may not occur on the path of the L\'evy process that is generated from the i.d. distribution with triplet $(b,\sigma^2,\nu)$. According to Theorem 26.8 of \cite{sato1999levy}, this condition requires that the i.d. variables $\xi_k$ should satisfy that $\mathbb{P}\{ \xi_k >t \} = o({\rm e}^{-\alpha t \log t})$ with $0\leq \alpha \leq R^{-1}$.




Note that the tail inequality \eqref{eq:tail2} is similar  to the matrix Bennett result ({\it cf.} \cite[Theorem 6.1]{tropp2012user}). Following the classical method of bounding $Q(s)$ from below, the Bernstein-type result can be derived based on the fact that 
\begin{equation}\label{eq:bernstein}
Q(s)\geq B(s)\geq
T(s),\quad s\geq 0,
\end{equation}
where 
\begin{equation}\label{eq:b.t}
B(s):= \frac{s^2}{2(1+s/3)};\;\; 
T(s):=\left\{
\begin{array}{cc}
 3s/4, & s\geq 3;      \\
   s^2/4, &  0< s <3. 
\end{array}
\right.
\end{equation}

\begin{figure}[htbp]
\centering
\subfigure[\hbox{$s\in(0,3]$}]{
\includegraphics[height=6cm]{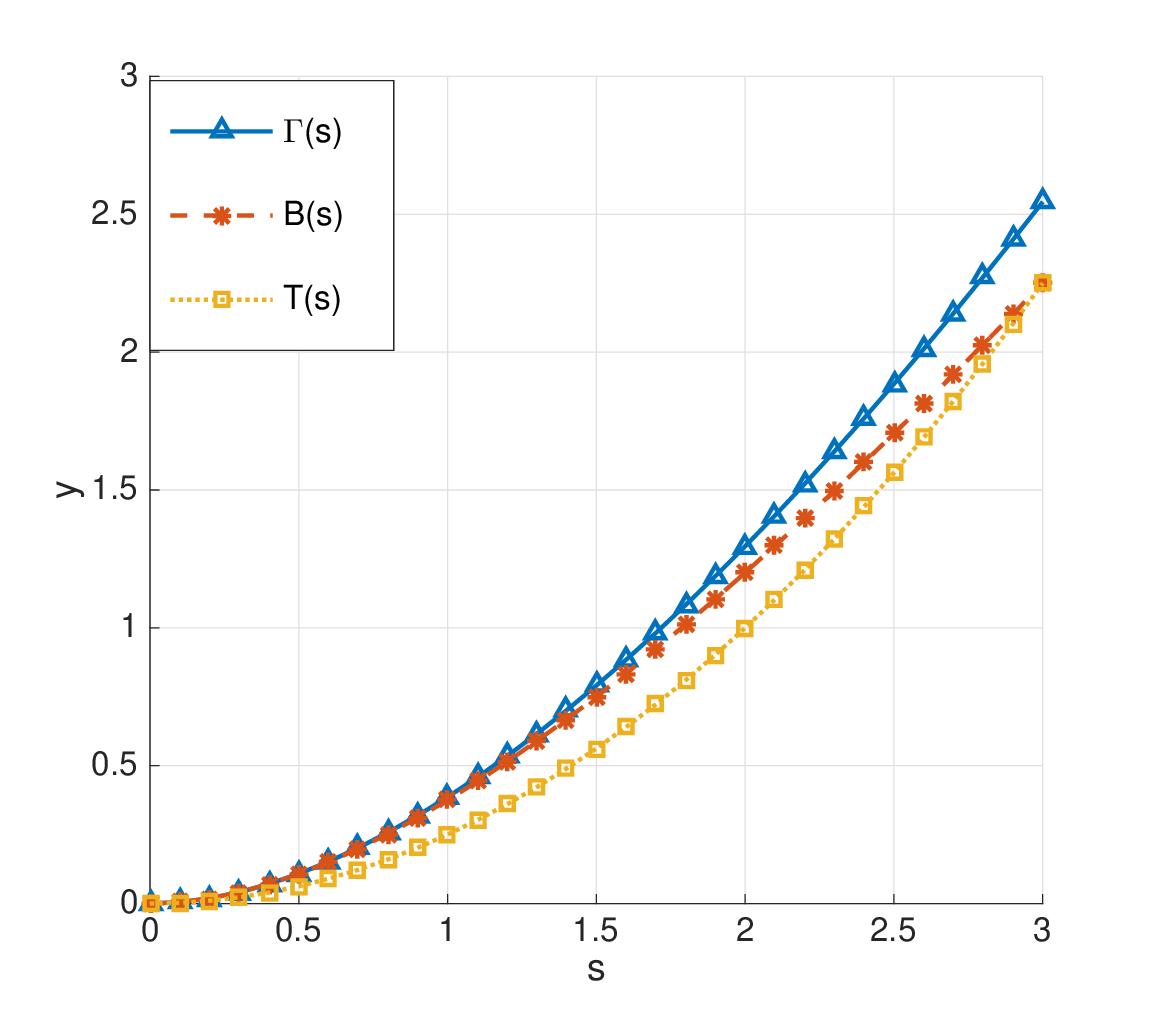} }
\subfigure[\hbox{$s\in(0,20]$}]{
\includegraphics[height=6cm]{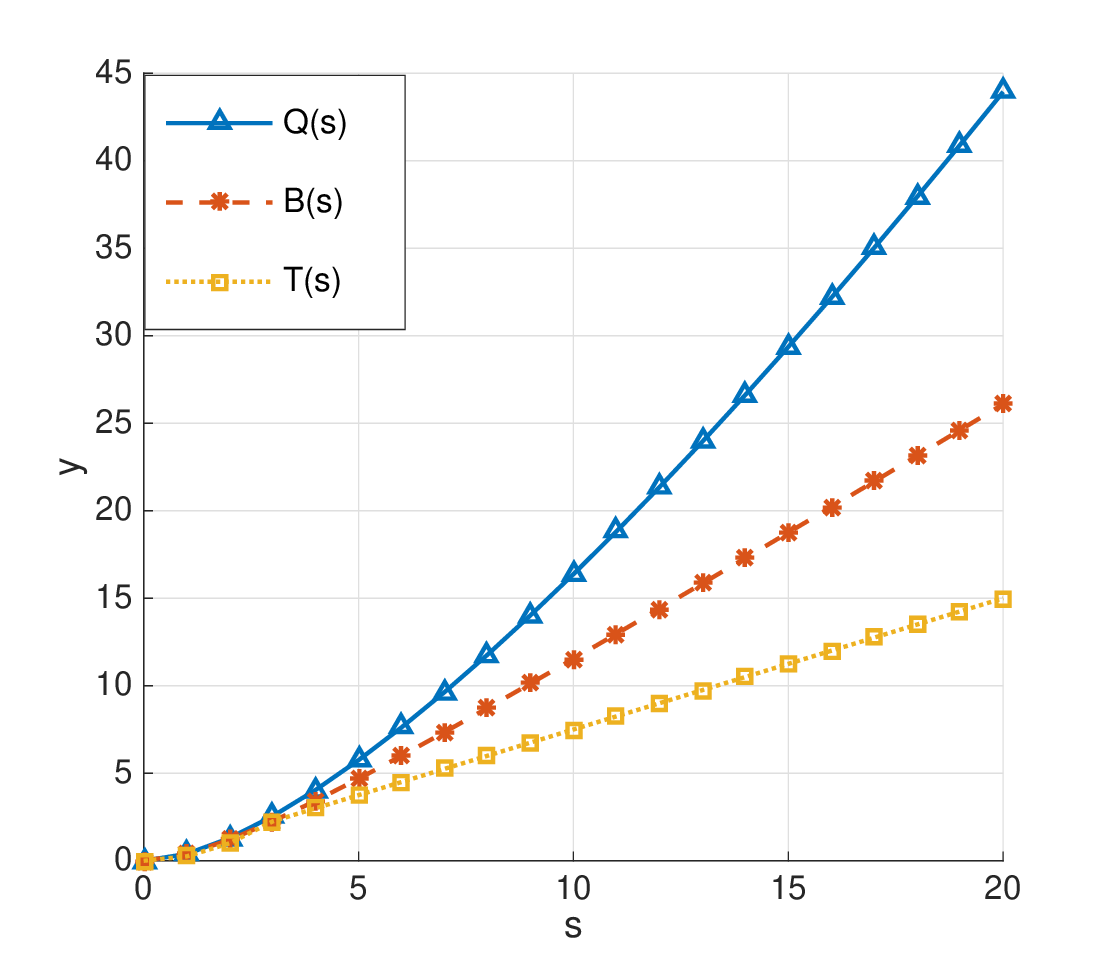}}
\caption{The function curves of $Q(s)$, $B(s)$ and $T(s)$.}\label{fig:QBT}
\end{figure}

As shown in Fig.~\ref{fig:QBT}, the function $B(s)=\frac{s^2}{2(1+s/3)}$ can tightly bound $Q(s)$ from below when $s$ is close to the {\it origin}, whereas there will be a large discrepancy between $Q(s)$ and $B(s)$ when $s$ is far from the {\it origin}. This is because $B(s)$ is derived from the Taylor expansion at the point $s=0$ ({\it cf.} \cite[Chapter 2.7]{boucheron2013concentration}). To facilitate the analysis of $Q(s)$, the function $B(s)$ is relaxed to a looser lower-bound function $T(s)$, which is a piecewise function with the following sub-functions: $s^2/4$ when $s\in(0,3)$; and $3s/4$ when $s\in[3,\infty)$. Although the function $T(s)$ does not bound $Q(s)$ sufficiently tightly, the result presented in \eqref{eq:beta1} below shows that $T(s)$ provides the same rate of growth as $Q(s)$ when $s$ is close to the {\it origin} or approaches {\it infinity}. This phenomenon suggests that the coefficients $3/4$ and $1/4$ of the sub-functions $3s/4$ and $s^2/4$, respectively, are probably not sufficiently well-tuned.

\begin{corollary}\label{cor:tail2}
Let $\xi_1,\cdots,\xi_K$ be independent i.d. random variables with bounded support $R$. 
Then for any $t>0$, 
\begin{align}
& \mathbb{P}\left\{\lambda_{\max}\left(\sum_k \xi_k {\bf A}_k\right)> t\right\} \nonumber\\
\leq&d\cdot \exp\left( - \frac{3}{2}\cdot \frac{t^2 }{3\rho(\sigma^2 +V) +  R t }  \right)\label{eq:tail3}\\
\leq&\left\{
  \begin{array}{ll}
 d\cdot \exp\left( -\frac{3}{4}\cdot\frac{t}{R}\right) , & \mbox{if $R t > 3\rho(\sigma^2 +V)$;} \\
d\cdot \exp\left( -\frac{t^2}{4\rho(\sigma^2+V)}\right) , & \mbox{if $0<R t \leq 3\rho(\sigma^2 +V)$.}
  \end{array}
\right.\label{eq:tail6}
\end{align}
\end{corollary}
This corollary shows that the probability of the event $\|\sum_k \xi_k {\bf A}_k \|> t$ is bounded by $O({\rm e}^{-c_1t})$ when $t$ is large and that its upper bound is of the form $O({\rm e}^{-c_2t^2})$ when $t$ is small. 
{Since the tail inequality \eqref{eq:tail2} incorporates the function $Q(s)=(1+s)\log(1+s) -s$ that also appears in Tropp's result \cite[Theorem 6.1]{tropp2012user}, our results share the same form of Tropp's matrix Bernstein inequalities in spite of the parameters $R$, $V$ and $\sigma^2$ that signify the characteristics of the i.d. random variables $\xi_1,\cdots,\xi_K$. }


Finally, we will derive  an upper bound on $\mathbb{E} \|\sum_k \xi_k {\bf A}_k \|$ for a random i.d. series whose proof follows from the tail bound presented in \eqref{eq:tail3}. Our result complements \cite[(4.9)]{tropp2012user}, where the expectation $\mathbb{E} \big\|\sum_k \xi_k {\bf A}_k \big\|$ for a random Gaussian series is bounded by the term $O[\sqrt{\log(c\cdot d)}]$. 

\begin{theorem}\label{thm:expectation}
Let $\xi_1,\cdots,\xi_K$ be independent i.d. random variables with bounded support {$R\geq3/4$}. Then
\begin{align}\label{eq:expection}
&\mathbb{E} \left\|\sum_k \xi_k {\bf A}_k  \right\| \leq \frac{4R}{3}\cdot\log\left( 2d\cdot {\rm e}^{1+\frac{ 9\rho^2(\sigma^2+V)^2 }{R^2}}  \right).
\end{align}
\end{theorem}
The proof of this theorem is given in Appendix~\ref{app_them2}.

The upper bound on $\mathbb{E} \|\sum_k \xi_k {\bf A}_k \|$ for a random i.d. series is of the form $O[\log(c\cdot d)]$, which differs from the Gaussian bound of $O[\sqrt{\log(c\cdot d)}]$. We note that the higher expectation bound for a matrix i.d. series arises from the introduction of the quantities $V$ and $R$ that control the behavior of the L\'{e}vy measure $\nu$. 

\begin{remark}\label{rem:sum}
Note that the aforementioned tail results for matrix i.d. series can be generalized to the scenario of sums of independent i.d. random matrices ${\bf X}_1,\cdots,{\bf X}_K$, all of whose entries are i.d. random variables with the generating triplet $(b,\sigma^2,\nu)$. As a starting point, we first obtain the mgf bound for the self-adjoint i.d. random matrix ${\bf X}$ with $(b,\sigma^2,\nu)$ and $\lambda_{\max}({\bf X}) \leq 1$: for all $0<\theta < M$,
\begin{equation}\label{eq:mgf3}
\mathbb{E} {\rm e}^{\theta  {\bf X}} \preceq {\rm e}^{\Phi(\theta)\cdot \mathbb{E}({\bf X}^2)},
\end{equation}
which can be proven in a manner similar to Lemma \ref{lem:id.mgf}. We then arrive at upper bounds on $\mathbb{P}\big\{\big\|\sum_k {\bf X}_k\big\|> t\big\}$ and $\mathbb{E} \big\|\sum_k {\bf X}_k  \big\|$ with the same forms as those of the proposed results for matrix i.d. series except that the term $\rho= \lambda_{\max}\big(\sum_k  {\bf A}_k^2\big)$ is replaced by $\rho_0= \lambda_{\max}\big(\sum_k  \mathbb{E}({\bf X}_k^2)\big)$ [{\it cf.} \eqref{eq:tail1}, \eqref{eq:tail2}, \eqref{eq:tail3}, \eqref{eq:tail4} and \eqref{eq:expection}]. These results can also be regarded as an extension of the existing vector-version results ({\it cf.} \cite{houdre2002remarks,zhang2011generalization}).

\end{remark}

 

\subsection{A Lower-Bound Function of $Q(s)$}

As discussed above, both $B(s)$ and $T(s)$ are lower bound functions for $Q(s)$, but they do not bound $Q(s)$ sufficiently tightly when $s$ is far from the {\it origin} ({\it cf.} Fig. \ref{fig:QBT}) because they stem from the Taylor expansion at the {\it origin}. We adopt a more direct strategy to analyze the behavior of the function $Q(s)$; for earlier discussions on this topic, refer to \cite{zhang2013risk,zhang2013bennett}.  

We consider the following inequality: for all $s>0$, 
\begin{equation}\label{eq:eq1}
  (s+1)\cdot\log(s+1)-s \geq \beta\cdot s^{\tau},
\end{equation}
where the parameter $\beta$ is expected to be a constant independent of $s$ such that $\beta\cdot s^{\tau}$ bounds $Q(s)$ from below as tightly as possible. For any $s\in(0,1)\cup(1,+\infty)$, define 
\begin{equation}\label{eq:eq2}
\tau(\beta,s):=\frac{\log\big((s+1)\log(s+1)-s\big)-\log(\beta)}{\log(s)}.
\end{equation} 
Then, it follows from L'Hospital's rule that, for all $\beta>0$, 
\begin{equation}\label{eq:beta1}
\lim_{s\rightarrow 0^+} \tau(\beta,s) = 2\;\;\mbox{and}\; \;\lim_{s\rightarrow +\infty} \tau(\beta,s) = 1. \end{equation}
The two limits in \eqref{eq:beta1} suggest that piecewise function $T(s)$ indeed captures the rate of growth of the function $Q(s)$ as $s$ approaches either the {\it origin} or {\it infinity}.

\begin{figure}[htbp]
\centering
\includegraphics[height=6cm]{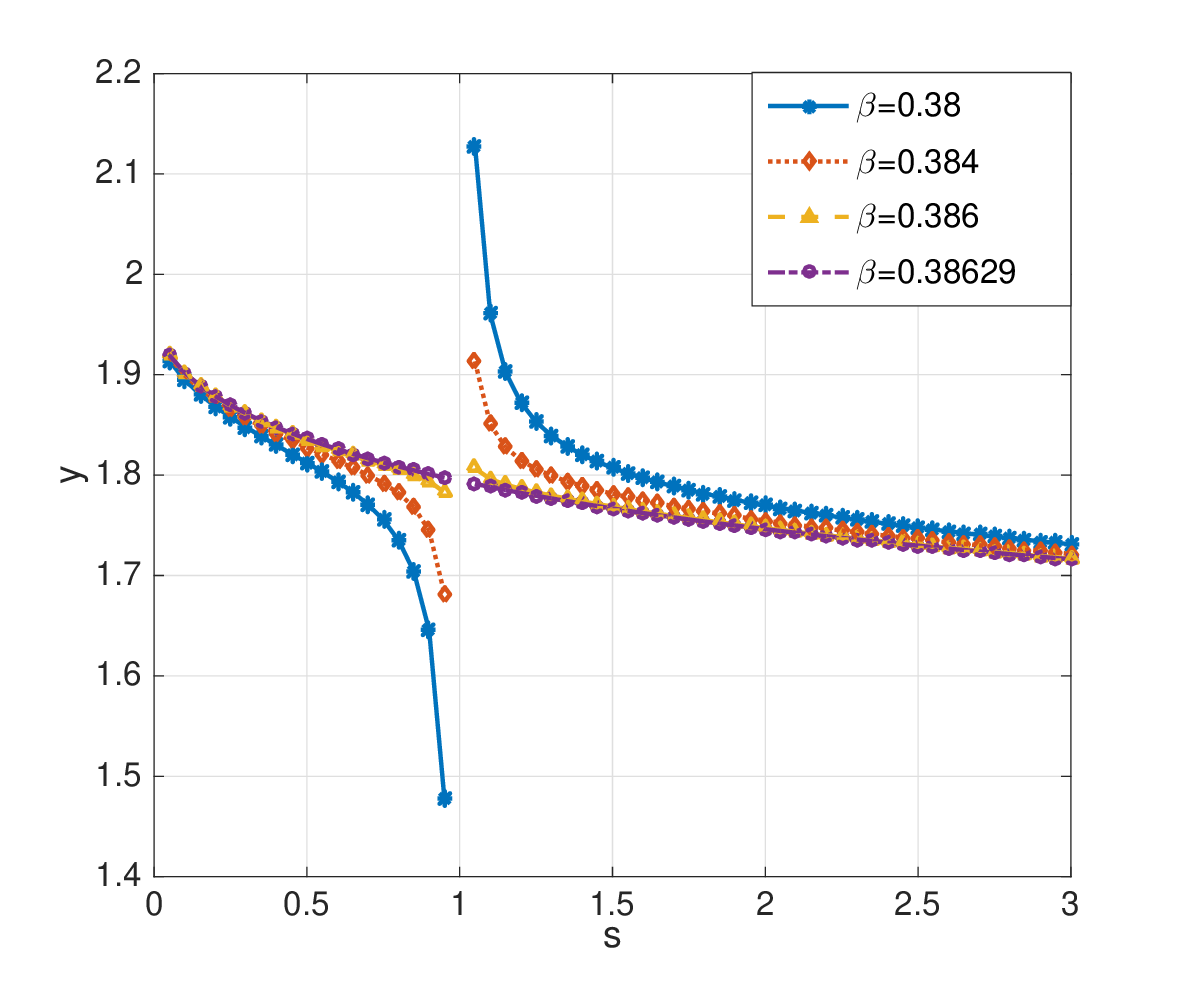}
\caption{The function curves of $\tau(\beta,s)$ w.r.t. different $\beta$ settings}\label{fig:beta}
\end{figure}

Now, we must choose the parameter $\beta$. As shown in Fig. \ref{fig:beta}, the function $\tau(\beta,s)$ is sensitive to the choice of $\beta$, and the value of $\tau(\beta,s)$ will vary dramatically near the point $s=1$ if parameter $\beta$ is not chosen well. Therefore, we should select a $\beta$ such that the variation of $\tau(\beta,s)$ near $s=1$ is kept as small as possible, {\it i.e.,} such that the discrepancy between $\tau(\beta,1^{-})$ and $\tau(\beta,1^+)$ is minimized. The follow lemma is also derived from L'Hospital's rule:
\begin{lemma}\label{lem:beta}
Let $\beta_0 = 2\log2-1$. Then, 
\begin{equation*}
\lim_{s\rightarrow 1^-}\tau(\beta_0,s)=\lim_{s\rightarrow 1^+}\tau(\beta_0,s)=\frac{\log2}{2\log2-1}.
\end{equation*}
\end{lemma}
This lemma shows that with the parameter choice $\beta = 2\log2-1$, the point $s=1$ is a removable discontinuity of the function $\tau(\beta,s)$; {\it i.e.,} $\tau(\beta,1^{-})=\tau(\beta,1^+)$. In other words, if we add a supplementary definition of $\tau(\beta,1):=\frac{\log2}{2\log2-1}$, the resulting function $\tau(\beta,s)$ will be continuous on the domain $(0,+\infty)$. Therefore, parameter $\beta$ should be selected such that $\beta=\beta_0=2\log2-1$.

%

By using the function $\tau(\beta_0,s)$, we can develop another lower-bound function for $Q(s)$ as follows.

\begin{proposition}\label{prop:H}
Given an arbitrary positive constant $c>1$ and an integer $N>1$, let $P
= \{p_0,p_1,\cdots,p_N\} $ be an ordered sequence such that $1=p_0<p_1<\cdots<p_{N-1}<p_N=c$, and define
\begin{equation}\label{eq:new1}
H_{P}(s):=
\left\{
\begin{array}{ll}
 \beta_0\cdot s^2, &0< s\leq p_0;      \\
   \beta_0\cdot  s^{\tau_1}, &   p_0<s\leq p_1;   \\
    \beta_0 \cdot s^{\tau_2}, &   p_1<s\leq p_2; \\
    \;\;\vdots& \qquad\vdots\\
    \beta_0\cdot  s^{\tau_n}, &   p_{N-1}<s\leq p_N,\end{array}
\right. 
\end{equation}
where $ \beta_0=2\log2-1$ and $\tau_n:=\tau(\beta_0,\,p_n)$ ($n=1,2,\cdots,N$). Then, for all $s\in(0,c]$, we have $Q(s)\geq H_{P}(s)\geq H_{\{1,c\}}(s)$, where the first equality holds when $s=p_0$ or $s=p_N$; and the second equality holds when $P = \{1,c\}$.

\end{proposition}

%

\begin{figure}[htbp]
\centering
\includegraphics[height=6cm]{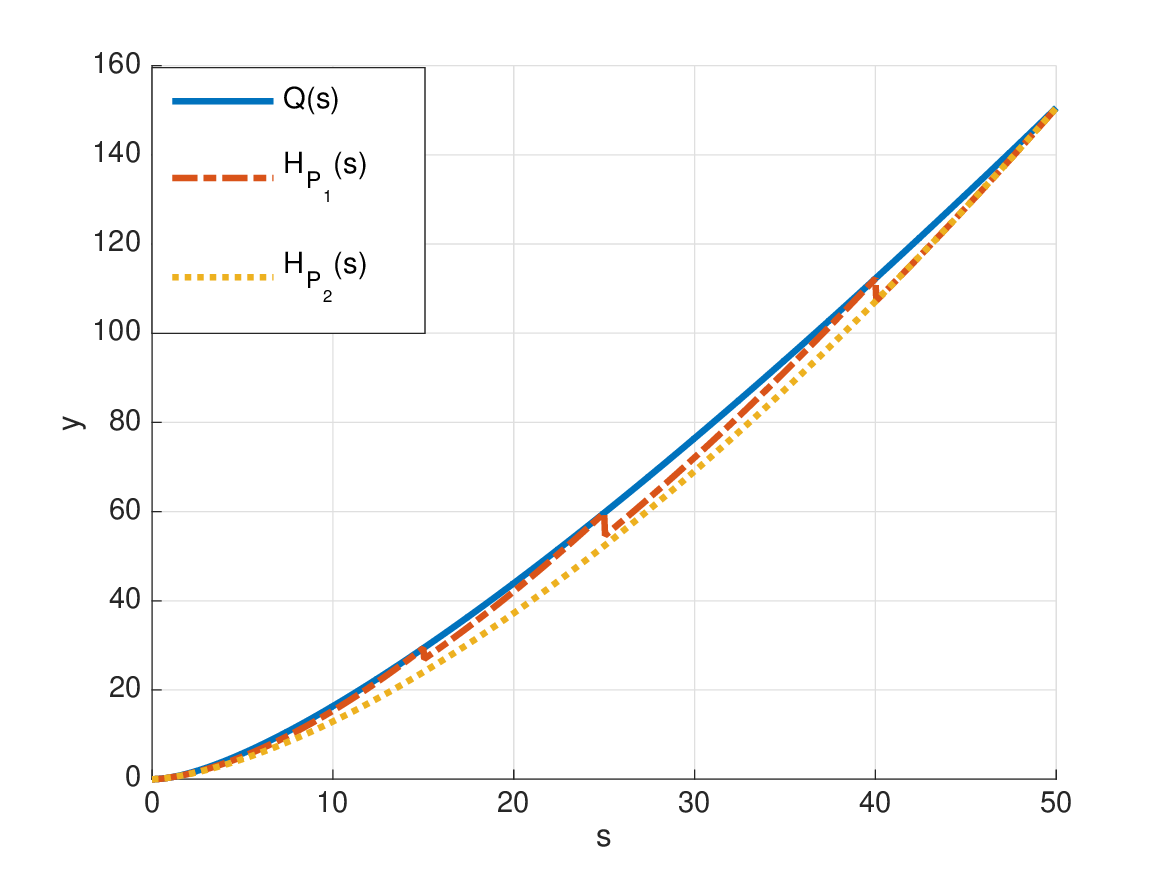}
\caption{The function curves of $H_{P}$ w.r.t. different $P$ settings, where $P_1 = \{1,   15  ,25  ,40  , 50\}$ and $P_2 = \{1,50\}$. Although the function $H_{P_1}$ is closer to $Q(s)$ than $H_{P_2}$ is, the curve of $H_{P_1}$ is not continuous and the discrepancy between $H_{P_1}$ and $H_{P_2}$ is not significant. }\label{fig:CompH}
\end{figure}

As suggested by this result, a piecewise function $H_{P}(s)$ to bound $Q(s)$ from below can be built when $s$ has a bounded domain $(0,c]$ by means of the following steps:
\begin{enumerate}[(i)]
\item Let $\beta_0 = 2\ln2-1$, and select a constant $c$ to form an interval $(0,c]$. 

\item Select an integer $N>1$ and an ordered sequence $P := \{p_0,p_1,\cdots,p_N\}$ such that $1=p_0<p_1<\cdots<p_{N}=c$.

\item If $s\in(0,1]$, then $H_{P}(s) = \beta_0s^2$; if $s\in (p_{n-1},p_n]$, then $H_{P}(s) = \beta_0s^{\tau_n}$, where $\tau_n = \tau(\beta_0,p_n)$ ($n=1,2,\cdots,N$).
\end{enumerate}
The resulting function $H_{P}(s)$ has the following characteristics:
\begin{itemize}

\item There is no additional restriction on the choice of the constant $c$, the integer $N$ and the points $p_1,p_2,\cdots,p_{N-1}$ other than $1=p_0<p_1<\cdots<p_{N}=c$. This means that suitable parameters $c$, $N$ and $\{p_1,p_2,\cdots,p_{N-1}\}$ can be chosen in accordance with the requirements of various practical problems. 

\item Although $H_{P}(s)$ is a piecewise function, all parts of $H_{P}(s)$ share the same coefficient $\beta_0=2\log2-1$, and the parameters $\tau_n$ are the values of function $\tau(\beta_0,s)$ at the partition points $p_n$ ($n=1,2,\cdots,N$). Therefore, the computation of $H_{P}(s)$ has a low cost.

\item For any choice of $P$, the piecewise function $H_{P}(s)$ has the same form $\beta_0\cdot s^2$ when $s\in(0,1]$. In particular, $H_{\{1,c\}}(s)$ ({\it i.e.}, with $P=\{1,c\}$) is a continuous function on $(0,c)$, and the difference between $H_{\{1,c\}}(s)$ and $H_{P}(s)$ is not significant for any other choice of $P$ ({\it cf.} Fig. \ref{fig:CompH}). Hence, $H_c(s):= H_{\{1,c\}}(s)$ can be adopted as the lower-bound function for $Q(s)$ if there are no additional requirements on the ordered sequence $P$.

\end{itemize}

\begin{figure}[htbp]
\centering
\subfigure[\hbox{$s\in(0,1]$}]{
\includegraphics[height=6cm]{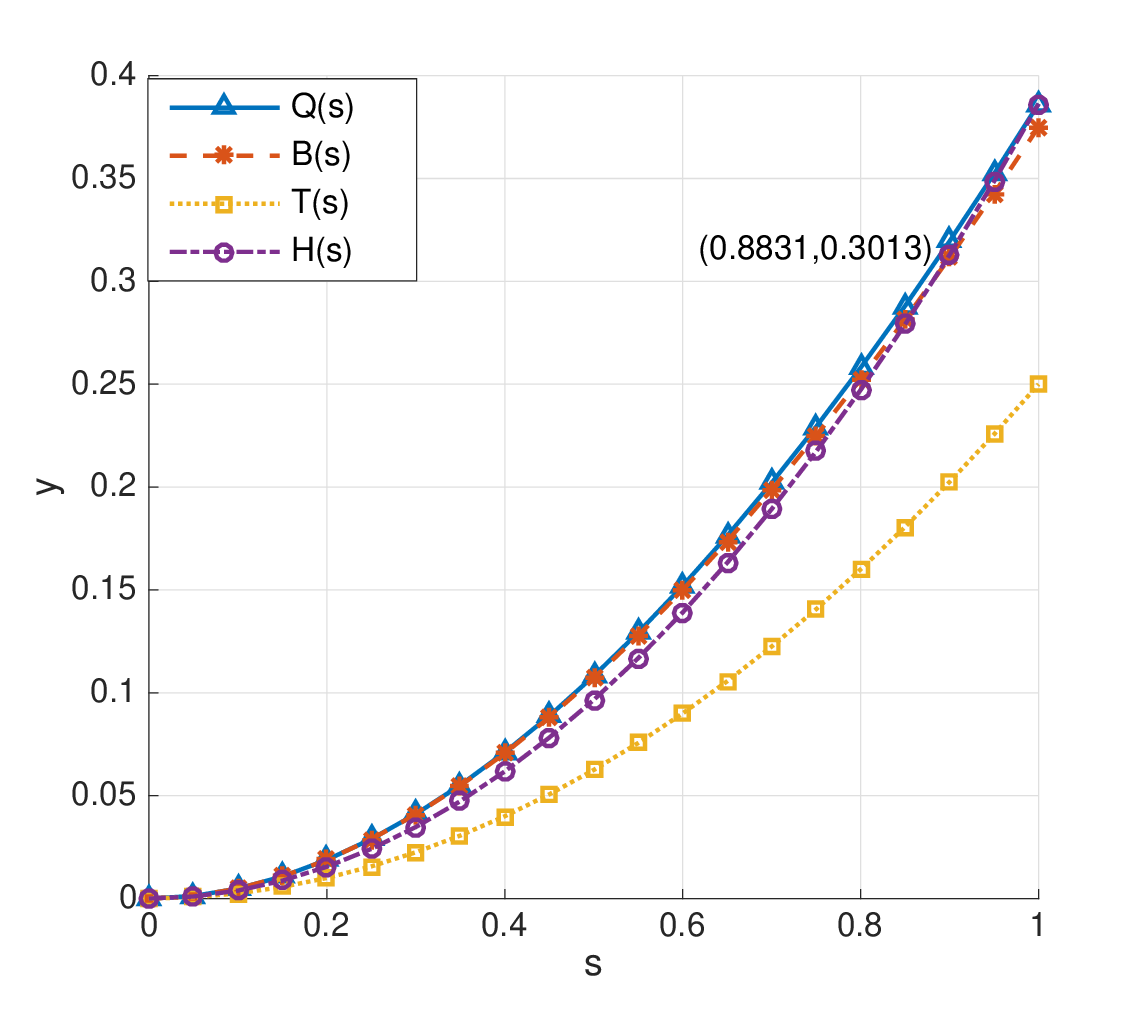} }
\subfigure[\hbox{$s\in(0,1000]$}]{
\includegraphics[height=6cm]{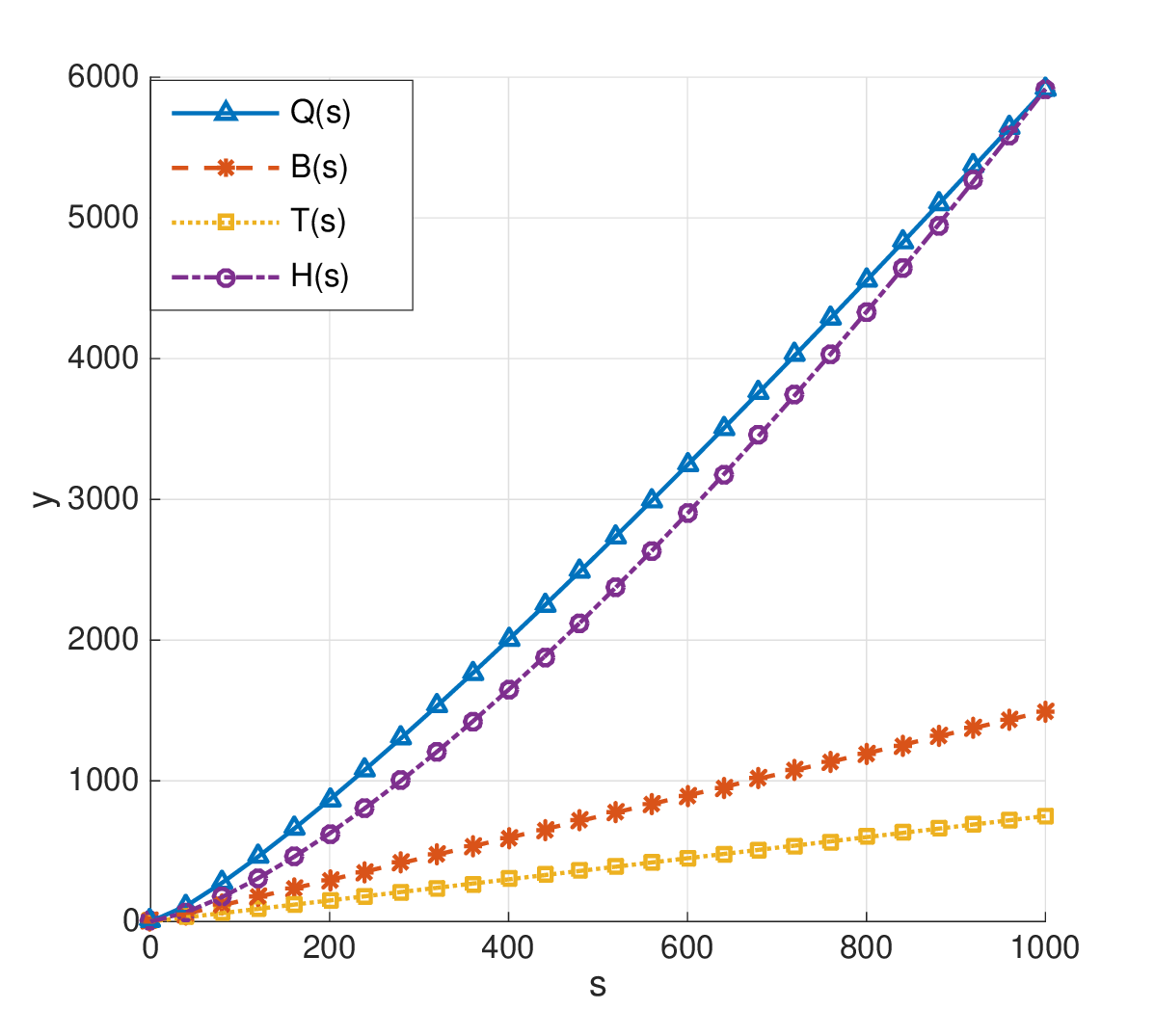}}
\caption{The function curves of $Q(s)$, $B(s)$, $T(s)$ and $H_c(s)$ with $c=1000$. The curves of $H_c(s)$ and $B(s)$ intersect approximately at the point $(0.8831,0.3013)$, and the function $H_c(s)$ is closer to $Q(s)$ than $B(s)$ is when $0.8831<s<1000$.}\label{fig:compare}
\end{figure}

\begin{remark}\label{rem:8831}

As shown in Fig. \ref{fig:compare}, the lower-bound function $H_c(s)$ performs better than the function $B(s)$, which is derived from the Taylor expansion, when $s\in(0.8831,c]$; moreover, although $B(s)$ bounds $Q(s)$ more tightly than $H_c(s)$ does when $s\in(0, 0.8831]$, there is only a slight discrepancy between $H_c(s)$ and $B(s)$ on this interval.\footnote{The range of $s\in(0.8831,c)$ is the numerical solution to the inequality $H_c(s)>B(s)$.} As a result, the method of bounding $Q(s)$ that is proposed in \eqref{eq:eq1} is not only effective but also corrects for the shortcoming of the Taylor-expansion-based method \eqref{eq:bernstein}, {\it i.e.,} the local approximation at the {\it origin}.

\end{remark}

\begin{figure}[htbp]
\centering
\includegraphics[height=6cm]{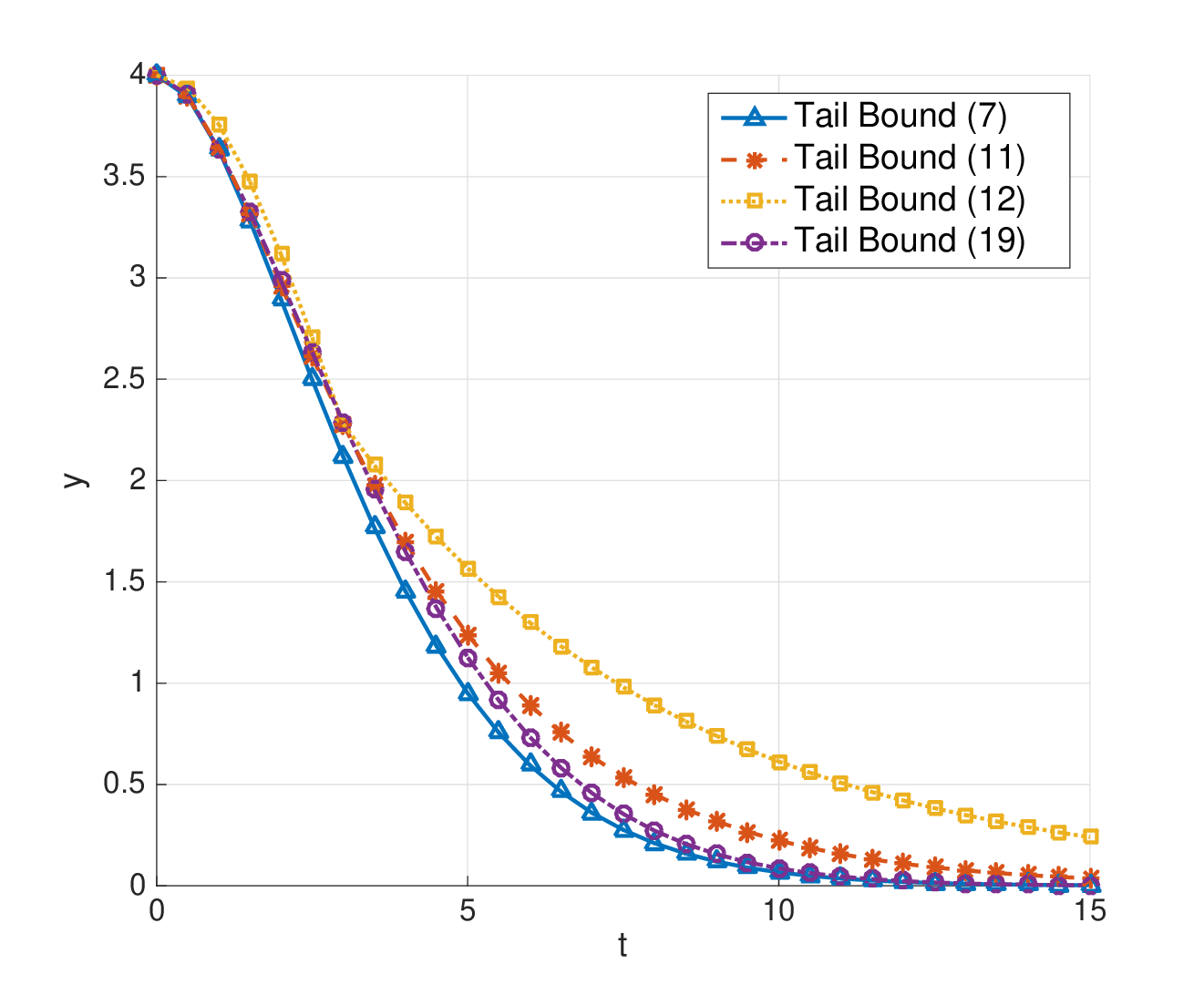}
\caption{The curves of the tail bounds \eqref{eq:tail2}, \eqref{eq:tail3}, \eqref{eq:tail6} and \eqref{eq:tail4}, where, for simplicity, the parameters are set as $d=2$, $R = 4$ and $\rho(\sigma^2+V)=4$.   }
\label{fig:bound_comp}
\end{figure}

By recalling the tail inequality \eqref{eq:tail2} and replacing the function $Q(s)$ with $H_c(s)$, we obtain, for any $0<\frac{R t }{\rho(\sigma^2 +V)}\leq c$,  
\begin{align}\label{eq:tail4}
& \mathbb{P}\left\{\lambda_{\max}\left(\sum_k \xi_k {\bf A}_k\right)> t\right\} \\
\leq&\left\{
  \begin{array}{ll}
d\cdot \exp\left( -\frac{\beta_0}{\rho(\sigma^2+V)}\cdot t^2\right) , & \mbox{if $0<\frac{Rt}{\rho(\sigma^2 +V)} \leq 1$;}\\
d\cdot \exp\left( - \frac{\beta_0\cdot R^{\tau_c-2}}{[\rho\cdot(\sigma^2+V)]^{\tau_c-1}}\cdot t^{\tau_c}\right), & \mbox{if $1<\frac{R t }{\rho(\sigma^2 +V)}\leq c$,} 
  \end{array}
\right.\nonumber
\end{align}
where $\tau_c = \tau(\beta_0,c)$. As shown in Fig. \ref{fig:bound_comp}, the above result provides a bound that is tighter than the one achieved by the Bernstein-type results in \eqref{eq:tail3} when $\frac{Rt}{\rho(\sigma^2+V)}\in (0.8831,c)$, and is only slightly looser than the Bernstein-type bound based on $B(s)$ when $\frac{Rt}{\rho(\sigma^2+V)}\in(0,0.8831)$.

\begin{remark}

Since the function $H_{c}(s)$ is defined on the bounded interval $(0,c]$, the result given in \eqref{eq:tail4} cannot be used to analyze the asymptotic behavior of $\mathbb{P}\{\lambda_{\max}(\sum_k \xi_k {\bf A}_k)> t\}$ as $t$ goes to {\it infinity}. However, since $H_{c}(s)$ bounds $Q(s)$ from below more tightly than $B(s)$ (or $T(s)$) does on the bounded domain $s\in(08831,c]$, the result given in \eqref{eq:tail4} provides a more accurate description of the non-asymptotic behavior of $\mathbb{P}\{\lambda_{\max}(\sum_k \xi_k {\bf A}_k)> t\}$ when $\frac{Rt}{\rho(\sigma^2+V)}>3$. The following alternative expressions for the Bernstein-type result given in \eqref{eq:tail3} and the $H_c$-based result given in \eqref{eq:tail4} can respectively be obtained: with probability at least $1-\delta$,
\begin{equation}\label{eq:alter.B}
\lambda_{\max}\left(\sum_k \xi_k {\bf A}_k\right)\leq \frac{4R (\log 2d - \log \delta)}{3}
\end{equation}
and 
\begin{equation*}
\lambda_{\max}\left(\sum_k \xi_k {\bf A}_k\right)\leq \left( \frac{\big(\log 2d - \log \delta\big) \big[\rho(\sigma^2+V)\big]^{\tau_c -1}}{ \beta_0 R^{\tau_c -2}  } \right)^{\frac{1}{\tau_c}}.
\end{equation*}
These expressions suggest that $\lambda_{\max}\big(\sum_k \xi_k {\bf A}_k\big)$ is bounded by the term $O(( \log d)^{\frac{1}{\tau_c}})$ with $1 < \tau_c <2$, which is a tighter bound than the right-hand side of the Bernstein-type result \eqref{eq:alter.B} when the matrix dimension $d$ is high.

\end{remark}


\section{Applications in Optimization}\label{sec:conjecture}

In this section, we will show that the derived tail inequalities for random i.d. series can be used to solve two types of optimization problems: chance constrained optimization problems and quadratic optimization problems with orthogonality constraints. These optimization problems are reviewed in Section~\ref{sec.app}, and Nemirovski's conjecture \cite{nemirovski2007sums} for efficiently computable solutions to these two optimization problems is introduced. We argue that the requirement in Nemirovski's conjecture is not practical, generalize the requirement using matrix i.d. series, and provide a solution to the extended version of Nemirovski's conjecture in Section~\ref{app.ext}. Lastly, we re-derive efficiently computable solutions to both types of optimization problems with a matrix i.d.~series requirement in Section~\ref{app.final}.


\subsection{Relevant Optimization Problems} \label{sec.app}

It has been pointed out in the pioneering work of \cite{nemirovski2007sums} that the behavior of $\sum_k \xi_k {\bf A}_k$ is  strongly related to the efficiently computable solutions to many optimization problems, {\it e.g.,} the chance constrained optimization problem and the quadratic optimization problem with orthogonality constraints. Several well-studied optimization problems are included in the latter as special cases, such as the Procrustes problem and the quadratic assignment problem. We begin with a brief introduction of these optimization problems.


\subsubsection{Chance Constrained Optimization Problem}

Consider the following chance constrained optimization problem ({\it cf.} \cite{so2011moment}): given   an $N$-dimensional vector ${\bf c}\in\mathbb{R}^N$  and an $\epsilon\in(0,1)$, find
\begin{align}\label{eq:chance}
&\min_{{\bf x}\in\mathbb{R}^N} {\bf c}^T {\bf x}\quad \mbox{subject to} \;\;  \\
&
\qquad\left\{
\begin{array}{ll}
 {\bf F}({\bf x})\leq {\bf 0}, & (a)   ;  \\
 \mathbb{P}\left\{ {\cal A}_0({\bf x}) - \sum_{k=1}^K \xi_k {\cal A}_k({\bf x}) \succeq {\bf 0}    \right\}\geq 1-\epsilon, &   (b),
\end{array}
\right.\nonumber
\end{align}
where ${\bf F}: \mathbb{R}^N \rightarrow \mathbb{R}^L$ is an efficiently computable vector-valued function with convex components; ${\cal A}_0,{\cal A}_1,\cdots,{\cal A}_K:\mathbb{R}^N \rightarrow \mathbb{S}^{M}$ are affine functions taking values in the space $\mathbb{S}^{M}$ of symmetric $M \times M$ matrices with ${\cal A}_0({\bf x}) \succeq {\bf 0}$ for all ${\bf x}\in\mathbb{R}^N$; and $\xi_1,\cdots,\xi_K$ are independent random variables with {\it zero} mean. The main challenge in solving this optimization lies in the chance constraint (\ref{eq:chance}-b). 

By letting ${\cal A}'_k ({\bf x})= ({\cal A}_0({\bf x}))^{-1/2}{\cal A}_k ({\bf x})({\cal A}_0({\bf x}))^{-1/2}$, we have
\begin{align*}
 \mathbb{P}\Big\{ {\cal A}_0({\bf x}) - \sum_{k=1}^K \xi_k {\cal A}_k({\bf x}) \succeq {\bf 0}    \Big\} = \mathbb{P}\Big\{\sum_{k=1}^K \xi_k {\cal A}'_k({\bf x}) \preceq {\bf I}    \Big\} .
\end{align*}
It is subsequently necessary to find a sufficient condition for the inequality 
\begin{align}\label{eq:chance0}
\mathbb{P}\left\{\sum_{k=1}^K \xi_k {\cal A}'_k({\bf x}) \preceq {\bf I}    \right\} \geq 1-\epsilon,
\end{align}
and to guarantee that the condition can be efficiently computable in optimization. For example, So proposed the following condition \cite{so2011moment}:
\begin{equation}\label{eq:chance1}
\sum_{k=1}^K ( {\cal A}'_k({\bf x}))^2 \preceq \gamma^2 {\bf I} \quad \mbox{with}\quad \gamma =\gamma(\epsilon) >0.
\end{equation}
By using the Schur complement, it can be equivalently expressed as a linear matrix inequality:
\begin{equation}\label{eq:chance2}
\left[\begin{array}{cccc}
  \gamma{\cal A}_0({\bf x})  &  {\cal A}_1({\bf x}) &  \cdots & {\cal A}_K({\bf x}) \\
 {\cal A}_1({\bf x})  & \gamma{\cal A}_0({\bf x})  &  & \\
 \vdots &   & \ddots  &\\
 {\cal A}_K({\bf x})  &   &   &\gamma{\cal A}_0({\bf x})
\end{array}
\right] \succeq {\bf 0}.
\end{equation}
If the constraint (\ref{eq:chance}-b) is replaced with the inequality \eqref{eq:chance2}, the chance-constrained optimization problem will become tractable. To guarantee the validity of this replacement, it is necessary to consider the following problem:
\begin{description}
\item[ (P1)] Is the condition \eqref{eq:chance1} sufficient for the inequality \eqref{eq:chance0}?
\end{description}

\subsubsection{Quadratic Optimization Problems with Orthogonality Constrains}\label{sec:qop}

Let $\mathbb{M}^{M\times N}$ be the space of $M\times N$ real matrices equipped with the trace inner product ${\bf X }\bullet {\bf Y} = {\rm tr}({\bf X }^T {\bf Y}) = {\rm tr}({\bf Y }^T {\bf X})$.
Consider the following quadratic optimization problem:
\begin{align}\label{eq:orthogonal0}
&\min_{{\bf X}\in \mathbb{M}^{M\times N}} {\bf X }\bullet {\cal A}{\bf X}\; \;\mbox{subject to}\nonumber\\
&\qquad\left\{\begin{array}{ll}
 {\bf X }\bullet \mathcal{B}_i{\bf X} \leq 1,\quad i=1,\cdots,I; &   (a)  \\
  \mathcal{C}{\bf X} = {\bf 0};&  (b)    \\
\|{\bf X}  \| \leq 1, &      (c)
\end{array}\right.
\end{align}
where ${\cal A},\mathcal{B}_1,\cdots,\mathcal{B}_I:\mathbb{M}^{M\times N} \rightarrow \mathbb{M}^{M\times N} $ are self-adjoint linear mappings (note that they can be represented as symmetric $MN\times MN$ matrices); $\mathcal{B}_1,\cdots,\mathcal{B}_I$ are positive semidefinite; $\mathcal{C} : \mathbb{M}^{M\times N}\rightarrow \mathbb{R}^L$ is a linear mapping (which can be represented as symmetric $L\times MN$ matrices); and $\|{\bf X}  \|$ is the spectral norm of ${\bf X}$. As addressed in \cite{nemirovski2007sums}, this optimization problem covers many well-studied optimization problems with the orthogonality constraint ${\bf X}^T{\bf X} = {\bf I}$ as special cases, {\it e.g.,} the Procrustes problem and the quadratic assignment problem. By exploiting the structure of these problems, the orthogonality constraint ${\bf X}^T{\bf X} = {\bf I}$ can be relaxed to the constraint (\ref{eq:orthogonal0}-c) without loss of generality.

The optimization problem can be directly tackled by using the semidefinite programming (SDP) relaxation:
\begin{align}\label{eq:orthogonal1}
&\min{\bf D}\bullet {\bf Y} \ \; \mbox{subject to}\notag\\
&\left\{
\begin{array}{ll}
{\bf B}_i \bullet {\bf Y} \leq 1,\quad  \;i=1,\cdots,I;&  (a)    \\
 {\bf C}^T{\bf C} \bullet  {\bf Y} = 0;&  (b)    \\
 {\cal S}({\bf Y} )\preceq {\bf I}_M,\;  {\cal T}({\bf Y}) \preceq {\bf I}_N;&      (c) \\
 {\bf Y}\in\mathbb{S}^{MN},\; {\bf Y} \succeq{\bf 0}, & (d)
\end{array}
\right.
\end{align}
where $\mathbb{S}^{MN}$ is the space of $MN\times MN$ symmetric matrices; ${\bf D},{\bf B}_1,\cdots,{\bf B}_I$ are the $MN\times MN$ symmetric matrices corresponding to the self-adjoint linear mappings ${\cal D},{\cal B}_1,\cdots,{\cal B}_I$ respectively; ${\bf C}$ is the $L \times MN$ matrix corresponding to the mappings ${\cal C}$; ${\cal S}: \mathbb{ S}^{MN} \rightarrow  \mathbb{ S}^{M}$ is the linear mapping such that given ${\bf X}\in\mathbb{M}^{M\times N}$, ${\bf X} {\bf X}^T \preceq {\bf I}_M$ if and only if ${\cal S}\big ( ({\rm Vec}\,{\bf X}) ( {\rm Vec}\,{\bf X} )^T\big) \preceq {\bf I}_M $; and ${\cal T}: \mathbb{ S}^{MN} \rightarrow  \mathbb{ S}^{N}$ is the linear mapping such that ${\bf X}^T {\bf X} \preceq {\bf I}_N$ if and only if ${\cal T}\big ( ({\rm Vec}\,{\bf X}) ( {\rm Vec}\,{\bf X} )^T\big) \preceq {\bf I}_N $. Refer to Section 3.1.1 of \cite{so2011moment} for details of these notations.


By using the ellipsoid method, the solution $\widehat{{\bf Y}}$ to the optimization problem \eqref{eq:orthogonal1} can be obtained with an additive error $\pi>0$ in polynomial time. That is, if $\theta^*$ is the optimal value of \eqref{eq:orthogonal1}, the ellipsoid method can be used for any $\pi>0$ to obtain a solution $\widehat{{\bf Y}}$ in polynomial time such that $\widehat{{\bf Y}}$ is feasible for \eqref{eq:orthogonal1} and satisfies $\widehat{\theta} := {\bf A} \bullet \widehat{\bf Y} \geq \theta^* - \pi$, where ${\bf A}$ is the $MN\times MN$ symmetric matrix corresponding to the self-adjoint linear mapping ${\cal A}$ in (\ref{eq:orthogonal0}). 

The solution $\widehat{{\bf X}}\in\mathbb{M}^{M\times N}$ to the optimization problem \eqref{eq:orthogonal0} can be achieved by using $\widehat{{\bf Y}}$ along with a degree of randomness. Since $\widehat{{\bf Y}}\succeq {\bf 0}$, there exists a positive semidefinite matrix $\widehat{{\bf Y}}^{1/2} \in \mathbb{S}^{MN}$ such that $\widehat{{\bf Y}} = \widehat{{\bf Y}}^{1/2}\widehat{{\bf Y}}^{1/2}$. Since $\widehat{{\bf Y}}^{1/2}{\bf A}\widehat{{\bf Y}}^{1/2}$ is also symmetric, it has a spectral decomposition $\widehat{{\bf Y}}^{1/2}{\bf A}\widehat{{\bf Y}}^{1/2} = {\bf U}^T \Lambda {\bf U}$, where $ {\bf U}$ is an $MN\times MN$ orthogonal matrix and $\Lambda$ is an $MN\times MN$ diagonal matrix. Let $\bm{\xi} = (\xi_1,\xi_2,\cdots,\xi_{MN})^T$ be an $M N$-dimensional random vector, where $\xi_n$ $(1\leq n\leq MN)$ are i.i.d. with {\it zero} mean and {\it unit} variance. The solution $\widehat{{\bf X}}$ is ultimately achieved via ${\rm Vec}\,\widehat{{\bf X}} =\widehat{{\bf Y}}^{1/2}{\bf U}^T \bm{\xi} $. Alternatively, $\widehat{{\bf X}}$ can be expressed as 
\begin{equation}\label{eq:hatx}
\widehat{{\bf X}} = \sum_{i=1}^{MN} \xi_i {\bf Q}_i, 
\end{equation}
where ${\bf Q}_i\in\mathbb{R}^{M\times N}$ and ${\rm Vec} \,{\bf Q}_i$ is the $i$-th column vector of the matrix $\widehat{{\bf Y}}^{1/2}{\bf U}^T$ ($1\leq i\leq MN$). To explore the quality of solution $\widehat{{\bf X}}$, the following problem should be considered:
\begin{description}
\item[(P2)] Does $\widehat{{\bf X}}$ act as a high-quality solution to the optimization problem \eqref{eq:orthogonal0} with a reasonable (at least larger than $1/2$) probability?
\end{description}

\subsection{An Extension of Nemirovski's Conjecture}\label{app.ext}

Nemirovski \cite{nemirovski2007sums} pointed out that the aforementioned two problems P1 and P2 can be reduced to a question about the behavior of the upper bound of ${\rm Pr }\{ \|\sum_k \xi_k {\bf A}_k \|>t\}$ and the ``optimal" answer to this question can be achieved by resolving the following conjecture:
\begin{conjecture}(\cite{nemirovski2007sums,so2011moment})\label{conjecture}
Let $\xi_1,\cdots,\xi_K$ be i.i.d. random variables with {\it zero} mean, each of which obeys either distribution supported on $[-1,1]$ or Gaussian distribution with {\it unit} variance. Let ${\bf A}_1,\cdots,{\bf A}_K$ be arbitrary $M\times N$ matrices satisfying 
\begin{equation*}
\sum_{k=1}^K {\bf A}_k {\bf A}_k^T  \preceq {\bf I}_M\quad \mbox{and} \quad \sum_{k=1}^K  {\bf A}^T_k {\bf A}_k  \preceq {\bf I}_N.
\end{equation*}
Then, whenever $t= O\big[\sqrt{\ln(M+N)}\big]$, we have
\begin{equation}\label{eq:conjecture}
\mathbb{P}\left\{    \left\|\sum_{k=1}^K  \xi_k {\bf A}_k \right\| > t\right\}\leq \theta_1\cdot \exp(-\theta_2\cdot t^{2}),
\end{equation}
where $\theta_1$ and $\theta_2$ are absolute constants.
\end{conjecture}

Nemirovski \cite{nemirovski2007sums} showed that the inequality \eqref{eq:conjecture} is achieved when $ t = O\big[(\ln(M+N))^{\frac{1}{6}}\big]$, while there is a gap between this value of $t$ and the conjectured value $ O\big[\sqrt{\ln(M+N)}\big]$. Anthony So used a non-commutative Khintchine inequality to show that  when $t= O\big[\sqrt{(1+\alpha)\ln\,\max\{M,N\}}\big]$, for any $\alpha\geq1/2$ ({\it cf.} \cite{so2011moment}),
\begin{equation}\label{eq:so}
\mathbb{P}\left\{    \left\|\sum_{k=1}^K  \xi_k {\bf A}_k \right\| > t\right\}\leq O\Big[(\max\{ M,N \})^{-\alpha}\Big].
\end{equation}

Note that these results are built under the assumption that $\xi_1,\cdots,\xi_K$ are  either Gaussian distributions or  distributions supported on $[-1,1]$. However, the assumption will not always be satisfied in practice. Therefore, we extend the content of the conjecture to the i.d. scenario, {\it i.e.,} whether the inequality \eqref{eq:conjecture} is still valid when $\xi_1,\cdots,\xi_K$ are independent i.d. random variables with {\it zero} mean and {\it unit} variance. The following theorem provides a solution to the extended version of Nemirovski's conjecture.

\begin{theorem}\label{thm:conjecture}
 Assume that ${\bf A}_1,\cdots,{\bf A}_K$ are fixed $M\times N$ matrices satisfying $\lambda_{\max}( \mathfrak{D}({\bf A}_k))\leq 1$ for any $1\leq k\leq K$ and denote $\rho_1:= \lambda_{\max} (\sum_k\mathfrak{D}^2({\bf A}_k))$, where
 \begin{equation*}
\mathfrak{D}({\bf A}) := \left[
\begin{array}{cc}
 {\bf 0} & {\bf A}  \\
 {\bf A}^* & {\bf 0}   
\end{array}
\right].
\end{equation*} 
Let $\xi_1,\cdots,\xi_K$ be independent i.d. random variables with the triplet $(b,\sigma^2,\nu)$, each of which has {\it zero} mean and {\it unit} variance. Suppose that $\nu$ has a bounded support with $R=\inf\{a>0:\nu(\{u:|u|>a\})=0\}$ and set $V:= \int_{\mathbb{R}}|u|^2 \nu(du)$. For any $\alpha>0$, denote
\begin{align*}
c_\alpha := \frac{(1+\alpha) \ln(M+N)}{\sqrt{\beta_0}}\cdot & \max\{1,\sqrt{R}\} \nonumber\\
& \times \max\Big\{1, \sqrt{\frac{\rho_1(\sigma^2+V)}{R}}\Big\}.
\end{align*}
Let $\tau_\alpha := \tau(\beta_0,c_\alpha)\in(1,2]$, where $\tau(\cdot)$ is defined in \eqref{eq:eq2} and $\beta_0 = 2\ln2-1$. Then, when 
\begin{align}\label{eq:t2}
t =& \left[\frac{(1+\alpha)\cdot[\rho_1 (\sigma^2+V)]^{\tau_\alpha-1}\cdot  \ln(M+N) }{\beta_0\cdot R^{\tau_\alpha-2}}     \right]^{\frac{1}{\tau_\alpha}} \nonumber\\
&>\frac{\sigma^2+V}{R},
\end{align}
it holds that
\begin{equation}\label{eq:conjecture.final}
\mathbb{P}\left\{    \left\|\sum_{k=1}^K  \xi_k {\bf A}_k \right\| > t\right\}\leq (M+N)^{-\alpha},\quad \alpha>0.
\end{equation}

\end{theorem}

This theorem shows that if $\xi_1,\cdots,\xi_K$ are i.d.~distributions, the probability that 
$ \|\sum_{k=1}^K  \xi_k {\bf A}_k \| > t$ can also be bounded by the term $(M+N)^{-\alpha}$ ($\alpha>0$) when $t= O\big[((1+\alpha)\ln\max\{M,N\})^{1/\tau}]$ $(1< \tau\leq 2)$. 
This solution is in accordance with So's solution \eqref{eq:so} to the original Nemirovski conjecture up to some constant. Therefore, the discussion in Section \ref{sec.app} is also valid in the setting of matrix i.d.~series.

\begin{remark}
According to the tail inequality \eqref{eq:tail4}, when 
\begin{equation*}
t = \sqrt{\frac{(\alpha+1) \cdot [\rho_1(\sigma^2+V)]\cdot \ln(M+N)}{\beta_0}}\leq \frac{\sigma^2+V}{R},
\end{equation*}
the result \eqref{eq:conjecture.final} still holds. However, to satisfy this condition, an assumption about the distribution of the i.d. random variable $\xi_k$ needs to be imposed, {\it i.e.,} the value of $R$ should be small enough. This will restrict the generality of the result, so we omit it here.  
\end{remark}

\subsection{Solutions to Problems P1\&P2} \label{app.final}

In this section, we will provide solutions to the aforementioned problems P1 and P2 in the i.d.~scenario. By using the tail inequality \eqref{eq:tail4}, we first arrive at the solution to Problem P1:

\begin{theorem}\label{thm:solution1}
Consider the chance constrained optimization problem \eqref{eq:chance}. Let $\xi_1,\cdots,\xi_K$ be independent i.d.~random variables satisfying the conditions in Theorem \ref{thm:conjecture}.  Denote $\rho_2: = \lambda_{\max}(\sum_k( {\cal A}'_k({\bf x}))^2)$. For any $\epsilon\in(0,1/2]$, let $c>1$ satisfy that 
\begin{equation}\label{eq:solution1.cd2}
2M \exp \left(  -\frac{c^2 \beta_0 \rho_2 (\sigma^2+V)}{R^2} \right)     \leq \epsilon.
\end{equation}
If it holds that
\begin{equation}\label{eq:solution1.cd1}
\sum_{k=1}^K ( {\cal A}'_k({\bf x}))^2 \preceq \gamma {\bf I}
\end{equation}
with
\begin{equation*}
\gamma\leq \gamma_2(\epsilon):= \left( \frac{ \beta_0 R^{\tau_c-2} }{   \big[\rho_2(\sigma^2 +V)\big]^{\tau_c-1}  \log(\frac{2M}{\epsilon}) }   \right)^{\frac{1}{\tau_c}},
\end{equation*}
then the positive semidefinite constraint \eqref{eq:chance2} is a  tractable approximation of the constraint (\ref{eq:chance}-b).

\end{theorem}

Note that since $\tau_c=\tau(\beta_0,c)$ takes value from the interval $(1,\frac{\log(2)}{2\log(2)-1}\big)$ when $c>1$, $\gamma_2(\epsilon) = O\big(\log( \frac{2M}{\epsilon}  )^{-1/\tau_c}\big)$ is smaller than the value $\gamma = O\big( \log( \frac{M}{\epsilon}  )^{-1/2}  \big)$ obtained in the scenario of either the distributions with $[-1,1]$ support or Gaussian distributions ({\it cf.}  \cite{so2011moment}) when the matrix size $M$ is large. 

Next, we consider the solution to Problem P2 in the matrix i.d. scenario. Consider the quadratic optimization problem \eqref{eq:orthogonal0}.  The following theorem proves the properties of the solution $\widehat{{\bf X}}= \sum_{i=1}^{MN} \xi_i {\bf Q}_i$ in \eqref{eq:hatx}.

\begin{theorem}\label{thm:solution2}
Following the notations in \eqref{eq:orthogonal0} and \eqref{eq:orthogonal1}. Let $\xi_1,\cdots,\xi_{MN}$ be independent i.d. random variables satisfying the conditions in Theorem \ref{thm:conjecture}. Then, it holds that
\begin{description}
\item[i)] $\mathbb{E}\big\{ \widehat{{\bf X}}\bullet \mathcal{D} \widehat{{\bf X}}  \big\}= \widehat{\theta} $;
\item [ii)]$\mathbb{E}\big\{ \widehat{{\bf X}}\bullet \mathcal{B}_i \widehat{{\bf X}}  \big\}\leq 1, \quad i=1,\cdots, I$;
\item [iii)]$\mathcal{C} \widehat{{\bf X}} = 0 $;
\item[iv)] $\mathbb{E}\big\{ \widehat{{\bf X}}\widehat{{\bf X}}^T  \big\}= {\bf I}_M$ and $\mathbb{E}\big\{ \widehat{{\bf X}}^T\widehat{{\bf X}}  \big\}= {\bf I}_N$.
\end{description}
\end{theorem}
Its proof is similar to the proof of Proposition 1 in \cite{so2011moment}, so we omit it here.

This theorem shows that the matrix i.d. series $\widehat{{\bf X}}= \sum_{j=1}^{MN} \xi_j {\bf Q}_j$ satisfies the constraints of the original optimization problem \eqref{eq:orthogonal0} when taking expectation. It remains to justify whether $\widehat{{\bf X}}$ can also satisfy the constraints (\ref{eq:orthogonal0}-a) and (\ref{eq:orthogonal0}-c) with reasonable probability (at least larger than $1/2$).

\begin{theorem}\label{thm:solution3}
Assume that $\xi_1,\cdots,\xi_{MN}$ are independent i.d. random variables satisfying the conditions in Theorem \ref{thm:conjecture}. Let ${\bf B}'_i  ={\bf U} \widehat{{\bf Y}}^{1/2} {\bf B}_i  \widehat{{\bf Y}}^{1/2} {\bf U}^T $ ($i=1,\cdots, I$) and denote by ${\rm col}_j [({\bf B}'_i)^{1/2}]$ the matrix whose $j$-th column is the $j$-th column of the matrix $({\bf B}'_i)^{1/2}$ and the other entries are all {\it zero} ($j=1,\cdots, MN$). Denote $\rho_3 := \lambda_{\max} (\sum_{j=1}^{MN} {\bf Q}_j^2) $ and $\rho_4^{(i)}:= \lambda_{\max} \big(\sum_{j=1}^{MN} ( {\rm col}_j[({\bf B}'_i)^{1/2}]  )^2\big) $. Then, with probability at least $1/2$, it holds that
\begin{equation}\label{eq:solution3.1}
\|  \widehat{{\bf X}}  \| \leq  \left[\frac{3[\rho_3(\sigma^2+V)]^{\tau_2-1}\cdot \ln(M+N) }{\beta_0\cdot R^{\tau_2-2}}     \right]^{\frac{1}{\tau_2}},
\end{equation}
and for any $1\leq i\leq I$
\begin{equation}\label{eq:solution3.2}
 \widehat{{\bf X}} \bullet \mathcal{B}_i\widehat{{\bf X}}  \leq  \left[\frac{3[\rho_4^{(i)}(\sigma^2+V)]^{\tau_2-1}\cdot \ln(M+N) }{\beta_0\cdot R^{\tau_2-2}}     \right]^{\frac{2}{\tau_2}}.
\end{equation}
\end{theorem}

This theorem implies that 
\begin{equation*}
\overline{{\bf X}} :=  \widehat{{\bf X}}\cdot\left[\frac{3[\rho_*(\sigma^2+V)]^{\tau_2-1}\cdot \ln(M+N) }{\beta_0\cdot R^{\tau_2-2}}     \right]^{\frac{-1}{\tau_2}}
\end{equation*}
is feasible to the quadratic optimization problem \eqref{eq:orthogonal0} with a probability larger than $1/2$, where $\rho_* = \max\{ \rho_3,\rho_4^{(1)},\rho_4^{(2)},\cdots, \rho_4^{(I)}\}$. It thus also provides a solution to Problem P2.

%
%

\section{Applications in Compressed Sensing}\label{sec:rip}

In this section, we apply the resulted tail bounds to verify that if an measurement matrix of compressed sensing can be expressed as a random i.d. series, it still satisfies the restricted isometry property (RIP). We first give a brief introduction of the RIP in compressed sensing, and then present the main theorem about the RIP for random i.d. series.


\subsection{Compressed Sensing and Restricted Isometry Property}

Let ${\bf x}^\star \in\mathbb{C}^D$ be a vector (or signal) that is expected to be recovered by solving the underdetermined linear equation:
\begin{equation}\label{eq:cs.linear}
{\bf y} = {\bf A} {\bf x}^\star,
\end{equation}
where ${\bf y}\in\mathbb{C}^M$ ($M \ll D$) and ${\bf A} \in  \mathbb{C}^{M\times D}$ are called the measurement vector and the measurement (or sensing) matrix, respectively. The basic linear-algebra results show that there could arise infinitely many solutions to this linear equation (at least for a full rank ${\bf A}$). By imposing the additional condition that ${\bf x}^\star$ is $S$-sparse, {\it i.e.,}
\begin{equation*}
\| {\bf x}^\star\|_0 : = {\rm supp} ({\bf x}^\star) \leq S,
\end{equation*}
the linear equation \eqref{eq:cs.linear} can be equivalently reformulated as an $\ell_0$-minimization problem
\begin{equation}\label{eq:cs.l0min}
\min_{{\bf x}\in\mathbb{C}^D} \| {\bf x}\|_0\quad \mbox{subject to}  \quad {\bf A} {\bf x} = {\bf y}.
\end{equation}
Since this optimization problem is NP hard, an efficient way to solve it is to consider its convex relaxation:
\begin{equation}\label{eq:cs.l1min}
\min_{{\bf x}\in\mathbb{C}^D} \| {\bf x}\|_1\quad \mbox{subject to}  \quad {\bf A} {\bf x} = {\bf y},
\end{equation}
which is an $\ell_1$-minimization problem and can be solved with efficient convex optimization methods. However, there also remains a theoretical issue about the validity of the relaxation, {\it i.e.,} whether the solution to the problem \eqref{eq:cs.l1min} coincides the one to the problem \eqref{eq:cs.l0min}. In the literature on compressed sensing, it has been shown that if the measurement matrix ${\bf A} \in  \mathbb{C}^{M\times D}$ satisfies the RIP, the recovery $\hat{\bf x}$ from the $\ell_1$-minimization \eqref{eq:cs.l1min} can approximate the true ${\bf x}^\star$ well.

\begin{definition}\label{def:rip}
Given a matrix ${\bf A}\in\mathbb{C}^{M\times D}$, for any $0\leq S\leq D$, the restricted isometry constant of order $S$ of ${\bf A}$ is defined as the smallest number $\delta_S:=\delta_S({\bf A})$ such that
\begin{equation}\label{eq:rip1}
(1-\delta_S) \| {\bf x}\|^2_2 \leq \|{\bf A} {\bf x}  \|^2_2 \leq (1+\delta_S) \| {\bf x}\|^2_2,
\end{equation}
for all $S$-sparse ${\bf x}\in\mathbb{C}^D$. Let $\delta\in(0,1)$, we say that the matrix ${\bf A}$ satisfies the restricted isometry property (RIP) of order $S$ with parameter $\delta$, shortly, ${\rm RIP}_S(\delta)$, if $0\leq \delta_S({\bf A}) <\delta$.
\end{definition}

For convenience of the following discussion, we also introduce an alternative definition of the RIP condition and refer to \cite{baraniuk2008simple} for its details. Given a matrix ${\bf A}\in\mathbb{C}^{M\times D}$ and any set $\mathcal{I}\subseteq \{1,2,\cdots,D\}$ of column indices, denote by $[{\bf A}]_\mathcal{I}$ the $M\times |\mathcal{I}| $ matrix composed of these columns, where $|\mathcal{I}|$ stands for the cardinality of the set $\mathcal{I}$. Similarly, for a vector ${\bf x}\in\mathbb{C}^D$, we denote ${\bf x}_\mathcal{I} $ as the $|\mathcal{I}|$-dimensional vector obtained by retaining only the entries in ${\bf x}$ corresponding to the column indices in $\mathcal{I}$. Under these notations, we say that a matrix ${\bf A}$ satisfies the ${\rm RIP}_S(\delta)$ if there exists a $\delta_S\in(0,1)$ such that
\begin{equation}\label{eq:rip2}
(1-\delta_S) \| {\bf x}_\mathcal{I}\|^2_2 \leq \|[{\bf A}]_\mathcal{I} {\bf x}_\mathcal{I}  \|^2_2 \leq (1+\delta_S) \| {\bf x}_\mathcal{I}\|^2_2,
\end{equation}
holds for all sets $\mathcal{I}$ with $|\mathcal{I}| \leq S$. The condition \eqref{eq:rip2} is equivalent to requiring that all eigenvalues of the Gram matrix $[{\bf A}]^*_\mathcal{I}[{\bf A}]_\mathcal{I}$ lie in the interval $[1-\delta_S,1+\delta_S]$, where the superscript $^*$ stands for the conjugate transpose.

Many types of measurement matrices have been proven to satisfy the RIP condition with high probability, {\it e.g.}, random Gaussian or Bernoulli matrices ({\it cf.} \cite{candes2006near,mendelson2008uniform}). In view of the fast algorithm via the fast Fourier transform (FFT) and smaller amount of randomness, the circulant and Toeplitz matrices with Gaussian or Bernoulli entries have been designed as the measurement matrices in compressed sensing \cite{rauhut2009circulant,haupt2010toeplitz}. As addressed in \cite{tropp2015introduction}, both of the random circulant and Toeplitz matrices can be expressed as the random Gaussian (or Bernoull) series, {\it i.e.,} the form of $\sum_{k=1}^K \xi_k {\bf A}_k $ with ${\bf A}_k$ being fixed matrices and $\xi_k$ being Gaussian or Bernoulli. {However, there is limited discussion on the application of random i.d. series in compressed sensing. Subsequently, we will prove that if a measurement matrix can be expressed as an i.d. random matrix, it also satisfies the RIP with a high probability.}

\subsection{RIP of Random I.D. Series}

Given a random i.d. sereis ${\bf A} = \sum_{k=1}^K\xi_k {\bf A}_k$, consider the Gram matrix  
\begin{equation}\label{eq_id_Gram}
{\bf A}^* {\bf A} =\left(\sum_{k=1}^K\xi_k {\bf A}^*_k\right)\left(\sum_{k=1}^K\xi_k {\bf A}_k\right) = \sum_{j,k =1 }^K \xi_j\xi_k {\bf A}_j^*{\bf A}_k.
\end{equation}
 Two issues arise in order to apply our main results in Section~\ref{sec:basic} to prove that a random i.d.~series ${\bf A}$ satisfies the ${\rm RIP}_S(\delta)$ condition, namely, all the eigenvalues of its Gram matrix (\ref{eq_id_Gram}) lie in the interval $[1-\delta_S,1+\delta_S]$. In oder to make sure that the Gram matrix (\ref{eq_id_Gram}) is a random i.d. seriers, we have to {1) decouple the dependence among $\xi_j\xi_k$ ($1\leq j,k\leq K$), and 2) guarantee the products $\xi_j\xi_k$ ($1\leq j,k\leq K$) still obey i.d. distributions.} By using the decoupling principle ({\it cf.} Lemma \ref{lem:decoupling}), we obtain the following proposition that solves the first question:

\begin{proposition}\label{prop:aw}
Let ${\bf A}_1,\cdots,{\bf A}_K \in \mathbb{C}^{M\times S}$ be fixed matrices and $\xi_1,\cdots,\xi_K$ be i.i.d. random variables. Let $\{\zeta_{jk}\}$ $(1\leq j<k\leq K)$ be a sequence of independent random variables such that $\zeta_{jk}$ obeys the same distribution of $\xi_j\xi_k$ for any $1\leq j<k\leq K$. Denote 
\begin{equation*}\label{eq:constructB}
{\bf B}_{jk}  = \left\{
\begin{array}{cc}
 4^{2K-3}\cdot({\bf A}_j^* {\bf A}_k + {\bf A}_k^* {\bf A}_j),  &  \mbox{if $K$ is even};   \\
 4^{2K-1}\cdot ({\bf A}_j^* {\bf A}_k+{\bf A}_k^* {\bf A}_j),  &  \mbox{if $K$ is odd}.
\end{array}
\right.
\end{equation*}
Then, we have 
\begin{multline}\label{eq:aw}
\mathbb{E} {\rm tr} \exp \left( \sum_{j,k=1}^K \theta \xi_j \xi_k {\bf A}_j^* {\bf A}_k  \right) \\
\leq  S\cdot \exp\left(  \lambda_{\max} \left( \sum_{1\leq j<k\leq K}\log \mathbb{E} \,{\rm e}^{\theta \zeta_{jk} {\bf B}_{jk}}\right)  \right).
\end{multline} 
\end{proposition}
In this proposition, we use the decoupling principle to obtain i.i.d. copies $\{\zeta_{jk}\}_{j<k}$ of $\{\xi_j\xi_k\}_{j<k}$ and meanwhile to eliminate the square terms $\xi_k^2$ $(1\leq k\leq K)$ ({\it cf.} the proof of Lemma \ref{lem:decoupling}).


 As addressed in \cite{bondesson2015class}, if independent variables $\xi_1,\xi_2$ obey a generalized gamma convolution (GGC) distribution, the product $\xi_1\xi_2$ still obeys a GGC distribution. Since GGC distributions belong to a subclass of i.d. distributions\footnote{As shown in \cite{james2008generalized}, if $\xi$ is an i.d. random variable with the triplet $(b,\sigma^2,\nu)$, then $\xi$ is said to obey a GGC distribution if $\xi$ takes value in $\mathbb{R}^+$ and the L\'evy measure can be expressed as $\nu (dx) = \frac{h(x)}{x}$ with $h(x) = \int_0^{+\infty} {\rm e}^{-xy}\mu(dy)$ for a $\sigma$-finite and positive measure $\mu$.}, if $\xi_1,\cdots,\xi_K$ obey a GGC distribution, it is feasible to signify the distribution of $\zeta_{jk}$ by using a generating triplet $(\bar{b},\bar{\sigma}^2,\bar{\nu})$ for all $1\leq j,k\leq K$ with $j\not=k$. The other thing to note is that since all GGC random variables takes values in $\mathbb{R}^+$, we will consider the centered random variable $\zeta - \mathbb{E}\zeta$ that actually has the triplet $(\bar{b}- \mathbb{E}\zeta, \bar{\sigma}^2,\bar{\nu} )$ and thus causes no effect to the final result because the term $\bar{b}- \mathbb{E}\zeta$ will be eliminated in Lemma \ref{lem:houdre2}. Denote
\begin{align}\label{eq:cs.notations}
\overline{R}:=&\inf\{a>0:\bar{\nu}(\{u:|u|>a\})=0\}; \nonumber\\
\overline{V}:=&\left| \int_{\mathbb{R}}|u|^2 \bar{\nu}(du)\right|.
\end{align}

\begin{lemma}\label{lem:cs}
Follow the notations in Proposition \ref{prop:aw}. Let ${\bf A} = \sum_{k=1}^K \xi_k{\bf A}_k $ be a random i.d. series where $\xi_1,\cdots,\xi_K$ obey a GGC distribution and ${\bf A}_1,\cdots,{\bf A}_K \in \mathbb{C}^{M\times S}$ are fixed matrices with $\lambda_{\max}({\bf B}_{jk})\leq 1$ ($1\leq j<k\leq K$). Denote $\bar{\rho} = \lambda_{\max}(\sum_{j<k}{\bf B}^2_{jk})$. Then, for any $0<\delta <1$, it holds that
\begin{equation}\label{eq:cs.lem1}
(1-\delta) \| {\bf x}  \|_2^2 \leq \| {\bf A}{\bf x}    \|_2^2 \leq (1+\delta) \| {\bf x} \|_2^2\quad  ({\bf x}\in \mathbb{R}^S)
\end{equation}
with probability at least 
\begin{equation*}
1- 2S\exp\left( - \frac{\bar{\rho}(\bar{\sigma}^2 +\overline{V})}{\overline{R}^2} \cdot Q \left(\frac{\overline{R}(1-\delta)}{\bar{\rho}(\overline{\sigma}^2+\overline{V})}\right) \right),
\end{equation*}
where $Q(s) = (s+1)\log(s+1)-s$.
\end{lemma}
This result suggests an upper bound of the probability that the matrix $[{\bf A}]_\mathcal{I}$ fails to satisfy the condition \eqref{eq:rip2} for any index set $\mathcal{I}\subset\{1,\cdots,D\}$ with $|\mathcal{I}| = S$. Subsequently, we will find the unified upper bound of the failure probability for all possibilities of the index set $\mathcal{I}$ and then multiply it by the combinatorial coefficient ${D \choose S}\leq ({\rm e} D/S)^S$ to achieve the upper bound of the probability that the ${\rm RIP}_S(\delta)$ fails to hold. It is noteworthy that the similar path of proof has been used in the earlier work \cite{baraniuk2008simple}, where the RIP is verified under the assumption that the measurement matrix should satisfy a concentration inequality relevant to the Johnson-Lindenstrauss lemma. In contrast, our resulted tail inequalities for random i.d. series have sufficiently provided a theoretical guarantee to the concentration behavior of the measurement matrix of interest. Therefore, the complexity of verifying RIP in this paper will not be higher than that in \cite{baraniuk2008simple}. 

\begin{remark}\label{rem:rho}
The following is a simple but rough way to bound the quantity $\bar{\rho}$ from above. Note that if ${\bf B}{\bf x} = \lambda {\bf x}$, then ${\bf B}^2{\bf x} = \lambda^2 {\bf x}$. It follows from the condition that $\lambda_{\max}({\bf B}_{jk})\leq 1$ that $\bar{\rho}\leq \sum_{j < k}\lambda_{\max}({\bf B}^2_{jk}) < {K \choose 2}$. 

\end{remark}

\begin{theorem}\label{thm:rip}
Let ${\bf A} = \sum_{k=1}^K \xi_k{\bf A}_k $ be a random i.d. series where $\xi_1,\cdots,\xi_K$ obey a GGC distribution and ${\bf A}_1,\cdots,{\bf A}_K \in \mathbb{C}^{M\times D}$ are fixed matrices. For any index set $\mathcal{I}\subset\{1,\cdots,D\}$ with $|\mathcal{I}| = S$, denote
\begin{align*}\label{eq:constructB}
{\bf B}^{\mathcal{I}}_{jk}  = \left\{
\begin{array}{ll}
 4^{2K-3}\cdot([{\bf A}_j]_\mathcal{I}^* [{\bf A}_k]_\mathcal{I} + [{\bf A}_k]^*_\mathcal{I} [{\bf A}_j]_\mathcal{I}),  &  \mbox{if $K$ is even};   \\
 4^{2K-1}\cdot ([{\bf A}_j]_\mathcal{I}^* [{\bf A}_k]_\mathcal{I} + [{\bf A}_k]^*_\mathcal{I} [{\bf A}_j]_\mathcal{I}),  &  \mbox{if $K$ is odd}.
\end{array}
\right.
\end{align*}
with $\lambda_{\max}({\bf B}^{\mathcal{I}}_{jk})\leq 1$ ($1\leq j<k\leq K$). Let 
\begin{equation}\label{eq:rho}
\bar{\rho}_S:= \max\limits_{\mathcal{I} \subset \{1,\cdots,D\}} \left\{ \lambda_{\max}\Big(\sum\limits_{j<k}({\bf B}^{\mathcal{I}}_{jk})^2\Big)\right\},
\end{equation}
Then, for any $1< \delta <0$, if there exists two positive constants $c_1,c_2$ such that 
\begin{equation}\label{eq:rip.cond1}
S \leq \frac{ c_1M    }{ \log({\rm e} D/S) },
\end{equation}
and
\begin{equation}\label{eq:rip.cond2}
c_2\leq \frac{1}{M}\left(\frac{\bar{\rho}_S(\bar{\sigma}^2 +\overline{V})}{\overline{R}^2} \cdot Q \left(\frac{\overline{R}(1-\delta)}{\bar{\rho}_S(\overline{\sigma}^2+\overline{V})}\right)\right) -c_1
\end{equation}
then the ${\rm RIP}_S(\delta)$ \eqref{eq:rip1} holds for the random i.d. series ${\bf A}$ with probability at least $1-2S\cdot{\rm e}^{-c_2 M}$.
\end{theorem}

As suggested in Remark \ref{rem:rho}, the quantity $\bar{\rho}_S$ has a simple but rough upper bound, {\it i.e.,} $\bar{\rho}_S< {K \choose 2}$, which implies that $\bar{\rho}_S$ could be related with the summand number $K$. Moreover, as required in Condition \eqref{eq:rip.cond2}, the validity of ${\rm RIP}_S(\delta)$ will be restricted by the term $S$. A large $S$, {\it i.e.,} the vector ${\bf x}$ has a low sparsity, possibly brings a low probability that ${\rm RIP}_S(\delta)$ holds. This finding reflects that the sparsity condition plays an essential role in compressed sensing. Recalling the process of proof, the appearance of $S$ is caused by the resulted tail bounds that have the matrix dimension as factor. Unfortunately, it is still an opening question about how to obtain the tail bounds for sums of random matrices without the matrix dimension as factor.


\section{Conclusion}\label{sec:con}

The class of i.d.~distributions is large and includes important probability distributions, such as Gaussian and Poisson distributions, that are widely used in several fields. To the best of our knowledge, however, little work has been done on random matrix theory with respect to i.d.~distributions. In this paper, we are mainly concerned with the tail inequalities of the largest eigenvalue of a matrix i.d.~series, and our results encompass Tropp's work \cite{tropp2012user} on matrix Gaussian series as a special case. Our proof strategy is as follows. We first relax the Bennett-type result \eqref{eq:tail2} into a Bernstein-type result \eqref{eq:tail3} by replacing $Q(s)$ with $B(s)$ or $T(s)$ \eqref{eq:bernstein}. Subsequently, we present an upper bound on the expectation $\mathbb{E}\big\|\sum_k \xi_k {\bf A}_k \big\| $, which is looser than the bound for the Gaussian case ({\it cf.} Inequality (4.9) of \cite{tropp2012user}) because of the existence of compound Poisson components in the i.d. distribution ({\it cf.} the L\'{e}vy-It\^{o} decomposition).

Since the function $B(s)$ does not bound $Q(s)$ from below sufficiently tightly ({\it cf.} Fig. \ref{fig:compare}), we develop a new lower-bound function $H_P(s)$ to bound $Q(s)$ from below on a bounded domain $s\in(0,c]$, where the partition $P = \{S_0,S_1,\cdots,S_N\}$ is an ordered sequence such that $1=S_0<S_1<\cdots<S_N=c$ for any given $c\in(1,+\infty)$. Although $H_P(s)$ is a piecewise function, its computational cost is low because all sub-functions of $H_P(s)$ are uniformly expressed in the form $\beta_0\cdot s^\tau_n$, where $\beta_0 = 2\log 2-1$ and $\tau_n = \tau(\beta_0,S_n)$ ($n=1,2,\cdots,N$). Based on $H_P(s)$, we obtain another tail inequality for matrix i.d. series that is tighter than the Bernstein-type result given in \eqref{eq:tail3} when $\frac{Rt}{\rho(\sigma^2+V)}>0.8831$ and provides a tighter upper bound on $\lambda_{\max}\big(\sum_k \xi_k {\bf A}_k\big)$ when the matrix dimension $d$ is high. Our results concerning the functions $Q(s)$ and $H_P(s)$ are also applicable for any Bennett-type concentration inequality that involves the function $Q(s)$.

In addition, we study the application of random i.d. series in several optimization problems including 1) the safe tractable approximation of chance constrained linear matrix inequalities, and 2) the quality of the semidefinite relaxation of a general non-convex quadratic optimization problem with orthogonality constraints, which covers two well-studied optimization problems as special cases: the Procrustes problem and the quadratic assignment problem. These two problems have been extensively studied in \cite{nemirovski2007sums,so2011moment} under the assumption that $\{\xi_k\}$ are sub-Gaussian, whereas in reality this assumption will not always be satisfied. We are able to extend the feasibility of the findings in \cite{nemirovski2007sums,so2011moment} to the case in which $\{\xi_k\}$ are i.d.~distributions. {Furthermore, we showed that if a measurement matrix in compressed sensing is constructed from generalized gamma convolution (GGC) distributions, then it satisfied restricted isometry property with high probability. 
 }

In order to achieve the tail results with well-defined forms [{\it e.g.} \eqref{eq:tail1} and \eqref{eq:tail2}], two conditions are imposed: one is that $M>0$ in Lemma \ref{lem:id.mgf} and the other is the L\'evy measure has a bounded support in Corollary \ref{cor:tail1}. The price we pay to obtain such results is to exclude some distributions from them, {\it e.g.,} heavy-tailed distributions. In the future work, we will consider the milder conditions to improve the generality of our results.

Since the tail inequalities considered in this paper depend on the matrix dimension, they will become loose in the high-dimensional case \cite{tropp2015introduction}. Similar to the results obtained in existing works, these inequalities can be improved by introducing the concept of effective dimension \cite{minsker2017on} or intrinsic dimension \cite{hsu2012tail}. In our future work, we will also consider the extension of these results to the infinite-dimensional case.


\appendices

\section{L\'evy Measure}\label{app:levy}

%

%



Before introducing the L\'{e}vy measure, we first present a discussion of L\'{e}vy processes. For further details, the reader is referred to {\cite{kyprianou2006introductory,sato1999levy,papapantoleon2008introduction,applebaum2009levy}}.
\begin{definition}[L\'evy Process]\label{def:levy}
A process $\mathcal{X} = \{X_t : t\geq0\}$, defined on a probability space $(\Omega,\mathcal{F},\mathbb{P})$, is said to be a L\'{e}vy process if it has the following properties:
\begin{enumerate}
\item The paths of $\mathcal{X}$ are $\mathbb{P}$-almost surely right continuous with left limits.
\item $\mathbb{P}(X_0=0)=1$.
\item For $0\leq s\leq t$, $X_t-X_s$ is equal in distribution to $X_{t-s}$.
\item For $0\leq s\leq t$, $X_t-X_s$ is independent of $\{X_u:u\leq s\}$.
\end{enumerate}
\end{definition}
Given a L\'{e}vy process $\{X_t: t\geq0\}$, consider the jump process $\Delta\mathcal{X} := \{{\Delta X}_t\}_{0\leq t\leq T}$, that is, for all $0\leq t\leq T$,
\begin{equation*}
\Delta X_t = X_t-X_{t^-}, 
\end{equation*}
where $X_{t^-}:=\lim_{s\rightarrow t^-}X_s$. It follows Definition \ref{def:levy} that
for any fixed $t > 0$, $\Delta X_t = 0$ almost surely. 

Moreover, given a set $A\in \mathcal{B}(\mathbb{R}/ \{0\})$ such that $0\notin \overline{A}$, let the random measure of the jumps be defined as 
\begin{align*}
\mu(\bm{\omega};t,A):=&\,\#\{0\leq s\leq t;\Delta X_s(\omega_s)\in A\}\\
=&
\sum_{s\leq t}1_A(\Delta X_s(\omega_s)),\qquad 0\leq t\leq T,\nonumber
\end{align*}
where $\bm{\omega}$ denotes joint probability events in the time interval $[0,t]$ and $\omega_s$ denotes events related to the $s$-time distribution of the L\'evy process $\{X_t: t\geq0\}$.
As defined above, the measure $\mu(\bm{\omega};t,A)$ counts the number of jumps of a size included in $A$ up to time $t$ in the process $\{X_t:t\geq 0\}$.

The L\'{e}vy measure is finally defined as
\begin{align*}
\nu(A): =& \mathbb{E}\big\{\mu(\bm{\omega};1,A)\big\}=\mathbb{E}\left\{\sum_{s\leq1}1_A(\Delta X_s(\omega_s))\right\},
\end{align*}
{which means that the} L\'evy measure {$\nu(A)$} describes the expected number of jumps of a certain height (belonging to $A$) in a time interval of {\it unit} length.



\section{Proofs of the Main Results}\label{app:proof}

Here, we prove Lemma \ref{lem:id.mgf}, Theorem \ref{thm:tail}, Corollary \ref{cor:tail1}, Theorem \ref{thm:expectation} and Theorem \ref{thm:conjecture}, Proposition \ref{prop:aw}, Lemma \ref{lem:cs} and and Theorem \ref{thm:rip}, respectively.

\subsection{Proof of Lemma \ref{lem:id.mgf}}\label{app.lem1}

Let $\psi(\theta):\mathbb{R}\rightarrow \mathbb{C}$ denote the characteristic function of the i.d.~random variable $\xi\in\mathbb{R}$ with the triplet $(b,\sigma^2,\nu)$. Let $(\xi_0,\xi'_0),(\xi_1,\xi'_1)\in\mathbb{R}\times\mathbb{R}$
be i.d. vectors\footnote{{A Borel probability measure $\mu$ of a random vector
$\bm{\xi}\in\mathbb{R}^K$ is infinitely divisible if and only if there exists a triplet
$({\bf b},\bm{\Sigma},\nu)$ such that
for all $\bm{\theta}\in\mathbb{R}^K$, its characteristic function
is of the form
\begin{multline*}
   \mathbb{E}\{ {\rm e}^{i \langle \bm{\theta},\bm{\xi}\rangle}\}=\exp\left( i\langle{\bf b},\theta\rangle-\frac{1}{2}\langle\theta,\bm{\Sigma}\theta\rangle\right. \\ \left.+\int_{\mathbb{R}^K\setminus\{0\}}\Big(\mathrm{e}^{i\langle\theta,u\rangle}-1-i\langle\theta,u\rangle{\bf 1}_{\|u\|\leq
  1}\Big)\nu(du)\right),
\end{multline*}
where ${\bf b} \in \mathbb{R}^K$, $\bm{\Sigma}$ is a $K\times K$
positive-definite symmetric matrix, and $\nu$ is a
L\'{e}vy measure on $\mathbb{R}^K\setminus\{0\}$.}} with the characteristic functions $\psi_0(\theta,\theta')=\psi(\theta)\cdot\psi(\theta')$ and $\psi_1(\theta,\theta')=\psi(\theta+\theta')$ $(\theta,\theta'\in\mathbb{R})$ respectively. For any $0\leq r\leq 1$, let $(\xi_r,\xi'_r)$ be a random vector with the characteristic function
\begin{align}\label{eq:psir}
   \psi_r(\theta,\theta'):=&\big[\psi_0(\theta,\theta')\big]^{1-r}\cdot \big[\psi_1(\theta,\theta')\big]^{r}\nonumber\\
   =&\big[\psi(\theta)\cdot \psi(\theta')\big]^{1-r}\big[\psi(\theta+\theta')\big]^{r}.
\end{align}

\begin{remark}
It is easy to see that $\psi_r(\theta,\theta')$ is a characteristic function. 
It follows from Theorem \ref{thm:cf} that $\big[\psi(\theta)\big]^r$ is the characteristic function of an i.d. random variable with the triplet $(rb,r\sigma^2,r\nu)$ and the fact that the product of a finite number of characteristic functions is also a characteristic function. 

\end{remark}

To prove Lemma \ref{lem:id.mgf}, we require the following two lemmas. The first one is the one-dimensional case of  \cite[Proposition 2]{houdre1998interpolation}.
\begin{lemma}\label{lem:houdre2}
Let $\xi$ be an i.d. random variable with the triplet $(b,\sigma^2,\nu)$. If $f,g:\mathbb{R}\rightarrow\mathbb{R}$ are differentiable functions such that $\mathbb{E}|f(\xi)|,\mathbb{E}|g(\xi)|,\mathbb{E}|f(\xi)g(\xi)|<\infty$, then
\begin{align*}
   &\mathbb{E}f(\xi)g(\xi)-\mathbb{E}f(\xi)\mathbb{E}g(\xi)
   =\int_0^1\mathbb{E}_r\left\{\sigma^2 \triangledown  f(\xi_r)\cdot \triangledown g(\xi_r)\right.\nonumber\\
   &\left.+\int_{\mathbb{R}}\big(f(\xi_r+u)-f(\xi_r)\big)\big(g(\xi'_r+u)-g(\xi'_r)\big)\nu(du)\right\}
   dr,
\end{align*}
where the expectation $\mathbb{E}_r$ is taken on the joint distribution of $(\xi_r,\xi'_r)$ and $\triangledown $ is the derivative notation.
\end{lemma}

{The second lemma below shows that for any $r\in[0,1]$, $\xi'_r$ and $\xi$ share the same characteristic function, which means that the distribution of $\xi'_r$ coincides with that of $\xi$.}
%
%
\begin{lemma}\label{lem:Er}
For any $r\in [0,1]$, it holds that
\begin{equation}\label{eq:Er}
{\psi_{\xi'_r}(\theta) = \psi(\theta)}.
\end{equation}
\end{lemma}
\begin{IEEEproof}[{Proof of Lemma \ref{lem:Er}}] According to \eqref{eq:psir}, for any $r\in[0,1]$, we arrive at
{\begin{align}\label{eq:Er.pr1}
\psi_{\xi_r'}(\theta) =& \mathbb{E}_r\big\{\mathrm{e}^{\theta \xi_r'}\big\} 
= \psi_r(0,\theta)\nonumber\\
=&\big[\psi( 0)\cdot \psi(\theta)\big]^{1-r}\big[\psi( 0+\theta)\big]^{r} \nonumber\\
=& \big[\psi( \theta)\big]^{1-r}\big[\psi(\theta)\big]^{r}
=  \psi(\theta) .
\end{align}}
%
This completes the proof. 
\end{IEEEproof}

Lemma \ref{lem:id.mgf} can be proven using the techniques presented in Houdr\'{e}'s work \cite{houdre2002remarks}.

\begin{IEEEproof}[{Proof of Lemma \ref{lem:id.mgf}}] As stated in Theorem 25.3 of \cite{sato1999levy}, since the function ${\rm e}^{ y }$ $(y\in \mathbb{R})$ is submultiplicative, it holds that
\begin{align*}
\Omega:=&\left\{s\geq 0 : \mathbb{E}\mathrm{e}^{s |\xi|}<+\infty\right\}\\
=&\left\{s \in\mathbb{R}: \int_{|u|>1}\mathrm{e}^{s|u|}\nu(du)<+\infty\right\}.
\end{align*}
Furthermore, it follows from the definition of the L\'evy measure $\nu$ ({\it cf.} Definition \ref{def:levy.meausre}) that
 \begin{align*}
\Omega=\left\{s\geq 0: \int_{|u|>1}\left(\mathrm{e}^{s|u|}-s |u|-1\right)\nu(du)<+\infty\right\},
\end{align*}
because $0< \mathrm{e}^{s|u|}-s |u|-1 < \mathrm{e}^{s|u|}$.
Based on the convexity of the exponential function, the set $\Omega$ is an interval of $\mathbb{R}$ and contains {\it zero}, but it cannot degenerate to $\{0\}$. We adopt the notation $\Omega = [0,M]$ with 
\begin{align*}
M=\sup\left\{s\geq 0: \int_{|u|>1}\mathrm{e}^{s|u|}\nu(du)<+\infty\right\}.
\end{align*}

By Lemma \ref{lem:houdre2}, we have
\begin{align}\label{eq:id.mgf1}
&\mathbb{E}\big\{\xi\cdot \mathrm{e}^{s\xi}\big\}
-\mathbb{E}\xi\cdot \mathbb{E}\mathrm{e}^{s\xi}\nonumber\\
=&\int_0^1\mathbb{E}_r\left\{\sigma^2\cdot\frac{{\rm d} \mathrm{e}^{s\xi'_r}}{{\rm d} \xi'_r} \right. \nonumber \\
&\quad \quad +\left.  \int_{\mathbb{R}}
(\xi_r+u-\xi_r)\big(\mathrm{e}^{s(\xi'_r+u)}-\mathrm{e}^{s\xi'_r}\big)\nu(du)\right\}dr\nonumber\\
= &\int_0^1\mathbb{E}_r\left\{s\mathrm{e}^{s\xi'_r}\sigma^2 +\mathrm{e}^{s\xi'_r}\int_{\mathbb{R}}
u\big(\mathrm{e}^{su}-1\big)\nu(du)\right\}dr\nonumber\\
\leq & \left(\sigma^2s+\int_{\mathbb{R}}|u|
\big(\mathrm{e}^{s |u|}-1\big)\nu(du)\right)\cdot\int_0^1\mathbb{E}_r\big\{\mathrm{e}^{s\xi_r'}\big\}dr\nonumber\\
=&\left(\sigma^2s+\int_{\mathbb{R}}|u|\big(\mathrm{e}^{s |u|}-1\big)\nu(du)\right)\cdot\mathbb{E}\big\{\mathrm{e}^{s\xi}\big\},
\end{align}
where the last equality follows from Lemma \ref{lem:Er}. 

Let $L(s):=\mathbb{E}\mathrm{e}^{s\xi'}$. It follows from $\mathbb{E}\xi = 0$ that
\begin{align*}
    \frac{{\rm d}L(s)}{{\rm d} s}\frac{1}{L(s)}=\frac{\mathbb{E}\xi\mathrm{e}^{s\xi}}{\mathbb{E}\mathrm{e}^{s\xi}}    \leq \sigma^2s+\int_{\mathbb{R}}|u| \left(\mathrm{e}^{s |u|}-1\right)\nu(du).
\end{align*}
Therefore, we have
\begin{align*}
  &\int_0^\theta \frac{{\rm d}L(s)}{{\rm d} s}\frac{1}{L(s)}ds\nonumber\\
  \leq& \int_0^\theta\left(\sigma^2s
  +\int_{\mathbb{R}} |u| \left(\mathrm{e}^{s|u| }-1\right)\nu(du)\right)ds,
\end{align*}
thus
\begin{align}\label{eq:id.mgf2}
\log\mathbb{E}\mathrm{e}^{s \xi}\Big|_{0}^\theta \leq\frac{\sigma^2 \theta^2}{2}
+\int_{\mathbb{R}}\left(\mathrm{e}^{\theta |u|}-\theta |u|-1\right)\nu(du).
\end{align}
From the proof of Lemma 6.7 in \cite{tropp2012user}, we obtain the following inequality: for any $ \lambda \leq 1 $
\begin{equation}\label{eq:id.mgf3}
\frac{\mathrm{e}^{\lambda\theta |u|}-\lambda\theta |u|-1}{\lambda^2} \leq \mathrm{e}^{ \theta|u|}-\theta|u|-1.
\end{equation}
Combining \eqref{eq:id.mgf2} and \eqref{eq:id.mgf3} yields, for any $\lambda \leq 1$, 
\begin{align*}
\mathbb{E}\mathrm{e}^{\lambda \theta \xi }
\leq \exp \left(\frac{\sigma^2 \theta^2\lambda^2}{2}
+\lambda^2\int_{\mathbb{R}}\left(\mathrm{e}^{ \theta|u|}- \theta|u|-1\right)\nu(du)\right).
\end{align*}
Given a self-adjoint matrix ${\bf A}$ with $\lambda_{\max}({\bf A})\leq 1$, it follows from the transfer rule that
\begin{equation}\label{eq:id.mgf5}
\mathbb{E} {\rm e}^{\xi \theta  {\bf A}} \preceq {\rm e}^{\Phi(\theta)\cdot{\bf A}^2},
\end{equation}
where 
\begin{equation}\label{eq:id.mgf6}
\Phi(\theta) := \frac{\sigma^2 \theta^2}{2}
+\int_{\mathbb{R}}\left(\mathrm{e}^{ \theta|u|}- \theta|u|-1\right)\nu(du),
\end{equation}
and $0<\theta < M$.
This completes the proof. 
\end{IEEEproof}


\subsection{Proof of Theorem \ref{thm:tail}}
\label{app.thm1}

\begin{IEEEproof}
Recall $\rho: = \lambda_{\max}\big(\sum_k  {\bf A}_k^2\big)$. It follows from Lemma \ref{lem:id.mgf} that, for any $t>0$,
\begin{align}\label{eq:pr.thmtail1}
&\mathbb{P}\left\{\lambda_{\max}\left(\sum_k \xi_k {\bf A}_k\right)> t\right\}\nonumber\\
\leq& {\rm e}^{-\theta t}\cdot {\rm tr} \exp\left(\sum_k \log \mathbb{E}{\rm e}^{\theta \xi_k {\bf A}_k}\right)\nonumber\\
\leq& {\rm e}^{-\theta t}\cdot {\rm tr} \exp\left(\Phi(\theta)\cdot \sum_k  {\bf A}_k^2\right)\nonumber\\
\leq& {\rm e}^{-\theta t}\cdot d \cdot \lambda_{\max}\left( \exp\left(\Phi(\theta)\cdot \sum_k  {\bf A}_k^2\right)\right)\nonumber\\
 = &  d \cdot \exp\left(-\theta t+\Phi(\theta)\cdot \lambda_{\max}\left(\sum_k  {\bf A}_k^2\right)\right)\nonumber\\
 = & d \cdot \exp\left(-\theta t+\Phi(\theta)\cdot\rho\right),  
\end{align}
where the first inequality follows from \cite[Theorem 3.6]{tropp2012user}.

Next, we minimize the right-hand side of \eqref{eq:pr.thmtail1} w.r.t. $\theta$. Since $\mathbb{E}\mathrm{e}^{\theta \xi}<+\infty$ for all $0<\theta<M$, $\Phi(\theta)$ in \eqref{eq:mgf2} is infinitely differentiable on $(0,M)$, with
\begin{align}\label{eq:Hn1}
   \Phi'(\theta)&:=\alpha(\theta)=\sigma^2\theta+\int_{\mathbb{R}}|u|
   \big(\mathrm{e}^{\theta |u|}-1\big)\nu(du)>0,
\end{align}
and
\begin{align}\label{eq:Hn2}
    \Phi''(\theta)=\sigma^2+\int_{\mathbb{R}}|u|^2\mathrm{e}^{\theta|u|}\nu(du)>0.
\end{align}
According to \eqref{eq:Hn1} and \eqref{eq:Hn2}, $\min_{0<\theta<M}\left\{\rho\cdot\Phi(\theta)-\theta\cdot t\right\}$ is achieved when $\theta=\alpha^{-1}(t/\rho)$ and $0<t<\rho\alpha(M^{-})$. Since $\Phi(0)=\alpha(0)=\alpha^{-1}(0)=0$, we have
\begin{align}\label{eq:Hn3}
    \Phi\left(\alpha^{-1}(t/\rho)\right)=&\int_{0}^{\alpha^{-1}(t/\rho)}\alpha(s)\,ds\nonumber\\
=&\int_{0}^{t/\rho}s\,d\alpha^{-1}(s)\nonumber\\
=&(t/\rho)\cdot\alpha^{-1}(t/\rho)-\int_{0}^{t/\rho}\alpha^{-1}(s)\,ds.
\end{align}
Thus, for any $0<t<\rho\alpha(M^{-})$, 
\begin{align*}
    \min_{0<\theta<M}\left\{\rho\cdot \Phi(\theta)-\theta\cdot t\right\}=&\rho\cdot \Phi\left(\alpha^{-1}(t/\rho)\right)-t \cdot \alpha^{-1}(t/\rho) \nonumber\\
     =&-\rho\cdot\int_{0}^{t/\rho}\alpha^{-1}(s)ds.
\end{align*}
This completes the proof. 
\end{IEEEproof}

\subsection{Proof of Corollary \ref{cor:tail1}}\label{app.corr1}

\begin{IEEEproof}
Since the support is ${\rm supp}(\nu)\subseteq[-R,R]$, it holds that $\mathbb{E}\mathrm{e}^{\theta |\xi|}<+\infty$ for any $\theta>0$. Thus, we have
\begin{align}\label{eq:pr.tail2.1}
    \alpha(\theta)=&\sigma^2\theta+\int_{\mathbb{R}}|u|
   \big(\mathrm{e}^{\theta |u|}-1\big)\nu(du)\nonumber\\
=&\sigma^2\theta+\int_{|u|\leq R}|u|^2\left(\sum_{k=1}^\infty\frac{\theta^k|u|^{k-1}}{k!}\right) \nu(du)\nonumber\\
\leq& \sigma^2\theta+\int_{|u|\leq R}|u|^2\left(\sum_{k=1}^\infty\frac{\theta^k R^{k-1}}{k!}\right) \nu(du)\nonumber\\
=&\sigma^2\theta+V\left(\frac{\mathrm{e}^{\theta R}-1}{R}\right)
\leq (\sigma^2+V)\left(\frac{\mathrm{e}^{\theta R}-1}{ R}\right).
\end{align}
{Denote $\beta(\theta):= (\sigma^2+V)\big(\frac{\mathrm{e}^{\theta R}-1}{ R}\big)$ with the inverse function $\beta^{-1}(s)=\frac{1}{R} \cdot\log \big(1 + \frac{Rs}{\sigma^2 +V}\big)$ $(s> 0)$. Since $\alpha(\theta)$ and $\beta(\theta)$ ($\theta>0$) are strictly increasing functions, their inverse functions satisfy the relation $\beta^{-1}(s)\leq \alpha^{-1}(s)$ for all $s> 0$. }
%
By combining \eqref{eq:tail1} and \eqref{eq:pr.tail2.1}, we obtain, for any $t>0$,
\begin{align*}
&\mathbb{P}\left\{\lambda_{\max}\left(\sum_k \xi_k {\bf A}_k\right)> t\right\}\nonumber\\
\leq & d\cdot \exp\left(-\rho\cdot\int_{0}^{t/\rho}\alpha^{-1}(s)ds\right)\\
\leq & d\cdot \exp\left(-\rho\cdot\int_{0}^{t/\rho} \frac{1}{R} \cdot\log \left(1 + \frac{Rs}{\sigma^2 +V} \right) ds\right) \\
= &d\cdot \exp\left( - \frac{\rho(\sigma^2 +V)}{R^2}\cdot Q \left(\frac{R t }{\rho(\sigma^2+V)}\right) \right), \end{align*}
where $Q(s) := (1+s)\cdot \log(1+s)-s$. This completes the proof.  


\end{IEEEproof}


\subsection{Proof of Theorem \ref{thm:expectation}}\label{app_them2}

\begin{IEEEproof}
 By combining Remark \ref{rem:spectral} and \eqref{eq:tail3}, we have
\begin{align}\label{eq:tail5}
&\mathbb{P}\left\{\left\|\sum_k \xi_k {\bf A}_k\right\|> t\right\}  \nonumber \\
&\leq{\left\{
  \begin{array}{ll}
 2d\cdot \exp\left( -\frac{3}{4}\cdot\frac{t}{R}\right) , & \mbox{if $t > \frac{3\rho(\sigma^2 +V)}{R}$;} \\
2d\cdot \exp\left( -\frac{t^2}{4\rho(\sigma^2+V)}\right) , & \mbox{if $0<t \leq \frac{3\rho(\sigma^2 +V)}{R}$.}
  \end{array}
\right.}
\end{align}
%
{Since $R\geq3/4$, it holds that $\frac{4R}{3} \cdot\log\big( 2d\cdot {\rm e}^{\frac{ 3\rho(\sigma^2+V) }{R}} \big) >  \frac{3\rho(\sigma^2 +V)}{R} $. Then, we have}
\begin{align}\label{eq:expection}
&\mathbb{E} \left\|\sum_k \xi_k {\bf A}_k \right\| \nonumber\\
= &\int_{0}^{+\infty}    \mathbb{P}\left\{\left\|\sum_k \xi_k {\bf A}_k \right\|> t \right\} d t       \nonumber\\
\leq & \frac{4R}{3} \cdot \log\left( 2d\cdot {\rm e}^{\frac{ 3\rho(\sigma^2+V) }{R}} \right) \nonumber \\ &\quad\quad\quad+2d\cdot \int_{\frac{4R}{3} \cdot\log\big( 2d\cdot {\rm e}^{\frac{ 3\rho(\sigma^2+V) }{R}} \big)}^{+\infty}  {\rm e}^{ -\frac{3t}{4R}} d \xi \nonumber\\
= &  \frac{4R}{3} \cdot \log\left( 2d\cdot {\rm e}^{\frac{ 3\rho(\sigma^2+V) }{R}} \right)  +  \frac{4R}{3} \cdot {\rm e}^{-\frac{ 3\rho(\sigma^2+V) }{R}} \nonumber\\
= & \frac{4R}{3} \cdot\log\left( 2d\cdot {\rm e}^{\frac{ 3\rho(\sigma^2+V) }{R}+{\rm e}^{-\frac{ 3\rho(\sigma^2+V) }{R}} }  \right) \nonumber\\
\leq &  \frac{4R}{3} \cdot\log\left( 2d\cdot {\rm e}^{1+\frac{ 9\rho^2(\sigma^2+V)^2 }{R^2}}  \right),\nonumber
\end{align}
{where the first equality is derived from the fact that $\mathbb{E}X =  \int_0^{+\infty} \mathbb{P}(X> x) dx$ holds for any non-negative random variable $X$, and the last inequality comes from the fact that $x+{\rm e}^{-x}\leq 1+x^2/2$ ($x> 0$).} This completes the proof. 
%
\end{IEEEproof}


\subsection{Proof of Theorem \ref{thm:conjecture}}

{\bf Proof of Theorem \ref{thm:conjecture}:} First, if $t$ satisfies the condition \eqref{eq:t2}, we have
\begin{align*}
t =& \left[(\alpha+1)\cdot \ln(M+N) \cdot \frac{[\rho_1(\sigma^2+V)]^{\tau_\alpha-1}}{\beta_0\cdot R^{\tau_\alpha-2}}     \right]^{\frac{1}{\tau_\alpha}}\nonumber\\
=&  \left[\frac{(\alpha+1)\cdot R\cdot \ln(M+N)}{\beta_0}    \right]^{\frac{1}{\tau_\alpha}}\cdot \frac{[\rho_1(\sigma^2+V)]^{1-\frac{1}{\tau_\alpha}}}{R^{1-\frac{1}{\tau_\alpha}}}. 
\end{align*}
Since it follows from \eqref{eq:eq2} and \eqref{eq:beta1} that $1<\tau_\alpha\leq2$ for any $\alpha>0$, we arrive at 
\begin{equation*}
t< \frac{(\alpha+1) \ln(M+N)}{\sqrt{\beta_0}}\cdot  \max\{1,\sqrt{R}\} \cdot \max\left\{1, \sqrt{\frac{\sigma^2+V}{R}}  \right\},
\end{equation*}
which suggests that $t< c_\alpha$ ($\alpha>0$). By using the dilation method ({\it cf.} Section 2.6 of \cite{tropp2012user}), we then have 
\begin{equation}\label{eq:conjecture.pr1}
\left\|  \sum_{k=1}^K \xi_k {\bf A}_k   \right\| = \lambda_{\max}\left(\sum_{k=1}^K \xi_k \mathfrak{D}({\bf A}_k) \right),
\end{equation}
where 
\begin{equation*}
\mathfrak{D}({\bf A}) := \left[
\begin{array}{cc}
 {\bf 0} & {\bf A}  \\
 {\bf A}^* & {\bf 0}   
\end{array}
\right].
\end{equation*}
Note that $s_{\max}({\bf A}_k)=\lambda_{\max}(\mathfrak{D}({\bf A}_k))\leq 1$ for all $k=1,2,\cdots, K$. According to \eqref{eq:tail4}, we then arrive at
\begin{align}\label{eq:conjecture.pr2}
&\mathbb{P}\left\{\left\|\sum_k \xi_k {\bf A}_k\right\|> t\right\} \nonumber\\
=& \mathbb{P}\left\{\lambda_{\max}\left(\sum_k \xi_k \mathfrak{D}({\bf A}_k)\right)> t\right\}\\
\leq&\left\{
  \begin{array}{l}
  (M+N)\cdot \exp\left( -\frac{\beta_0}{\rho_1(\sigma^2+V)}\cdot t^2\right), \\ \qquad\qquad\qquad\qquad\qquad\qquad\mbox{if $0<\frac{Rt}{\rho_1(\sigma^2 +V)} \leq 1$;}\\
   (M+N)\cdot \exp\left( - \frac{\beta_0\cdot R^{\tau_{c_\alpha}-2}}{[\rho_1(\sigma^2+V)]^{\tau_{c_\alpha}-1}}\cdot t^{\tau_{c_\alpha}}\right), \\
   \qquad\qquad\qquad\qquad\qquad\qquad \mbox{if $1<\frac{R t }{\rho_1(\sigma^2 +V)}\leq c_\alpha$,} 
  \end{array}
\right.\nonumber
\end{align}
Substituting \eqref{eq:t2} into the last inequality of \eqref{eq:conjecture.pr2} leads to the result \eqref{eq:conjecture.final}. This completes the proof. \hfill$\blacksquare$


\subsection{Proof of Theorem \ref{thm:solution1}}

{\bf Proof of Theorem \ref{thm:solution1}:} According to \eqref{eq:solution1.cd1}, it holds that $\lambda_{\max} \big( {\cal A}'_k({\bf x}) / \gamma \big)\leq 1$. We will consider two cases respectively: 1) $\gamma \geq \frac{R}{\rho_2(\sigma^2+V)}$; and 2) $\frac{R}{c\rho_2 (\sigma^2+V)} \leq \gamma< \frac{R}{\rho_2 (\sigma^2+V)}$ for an arbitrary $c>1$. 

When $\gamma \geq \frac{R}{\rho(\sigma^2+V)}$, it follows from \eqref{eq:tail2} that 
\begin{align}\label{eq:solution1.pr1}
&\mathbb{P}\left\{ \left\| \sum_{k} \xi_k \Big( \frac{1}{\gamma} {\cal A}'_k({\bf x}) \Big)    \right\| > \frac{1}{\gamma}\right\}\nonumber\\
\leq & 2M\exp\left\{ -\frac{\beta_0}{\rho_2(\sigma^2 +V) \gamma^2}  \right\}.
\end{align}
Given an $\epsilon\in(0,1/2)$, if it satisfies that $2M\exp\big\{ -\frac{\beta_0}{\rho(\sigma^2 +V) \gamma^2}  \big\}\leq \epsilon$, then the choice of $\gamma$ should satisfy that 
\begin{equation}\label{eq:solution1.pr2}
\gamma\leq\gamma_1(\epsilon):= \sqrt{\frac{\beta_0 }{\rho_2 (\sigma^2+V)\log\big(\frac{2M}{\epsilon} \big)}  },
\end{equation}
and meanwhile guarantee that $\frac{R}{\rho_2(\sigma^2+V)} \leq \gamma_1(\epsilon)$, which means that 
\begin{equation*}
\epsilon > 2M\cdot \exp\left( -\frac{\beta_0 \rho_2 (\sigma^2 + V)}{R^2}   \right).
\end{equation*}
This relation is only valid when $R$ is sufficiently large, so the case of $\gamma \geq \frac{R}{\rho_2(\sigma^2+V)}$ is not friendly enough to facilitate the optimization problem. We will omit this case

When $\frac{R}{c\rho_2 (\sigma^2+V)} \leq \gamma< \frac{R}{\rho_2 (\sigma^2+V)}$ for an arbitrary $c>1$, it also follows from \eqref{eq:tail4} that 
\begin{align}\label{eq:solution1.pr3}
&\mathbb{P}\left\{ \left\| \sum_{k} \xi_k \Big( \frac{1}{\gamma} {\cal A}'_k({\bf x}) \Big)    \right\| > \frac{1}{\gamma}\right\}\nonumber\\
\leq & 2M\exp\left\{ -\frac{\beta_0R^{\tau_c-2} }{\big[\rho_2(\sigma^2 +V)\big]^{\tau_c-1} \gamma^{\tau_c}}  \right\}.
\end{align}
For any $\epsilon\in(0,1/2)$, if the right-hand side of \eqref{eq:solution1.pr3} can be bounded by the constant $\epsilon$, the choice of $\gamma$ should satisfy the following condition:
\begin{equation*}
\gamma\leq \gamma_2(\epsilon):= \left( \frac{ \beta_0 R^{\tau_c-2} }{   \big[\rho_2(\sigma^2 +V)\big]^{\tau_c-1}  \log(\frac{2M}{\epsilon}) }   \right)^{\frac{1}{\tau_c}}.
\end{equation*}
It is clear that when 
\begin{align}\label{eq:eq:solution1.pr4}
     2M &\exp \left(  -\frac{c^2 \beta_0 \rho_2 (\sigma^2+V)}{R^2} \right)\nonumber\\
     &\leq \epsilon \leq 2M\exp\left(-\frac{\beta_0 \rho_2 (\sigma^2+V)}{R^2}\right),
\end{align}
it holds that $\frac{R}{c\rho_2 (\sigma^2+V)} \leq \gamma\leq \gamma_2(\epsilon)< \frac{R}{\rho_2 (\sigma^2+V)}$. The first inequality of \eqref{eq:eq:solution1.pr4} holds by setting appropriate $c>1$ and the second inequality holds when $\epsilon$ is small enough. Therefore, the validity of the inequality \eqref{eq:eq:solution1.pr4} is guaranteed. We then arrive at 
\begin{align}\label{eq:solution1.pr5}
&\mathbb{P}\left\{ \sum_{k} \xi_k {\cal A}'_k({\bf x})    \preceq {\bf I}\right\} \nonumber\\
= &
\mathbb{P}\left\{ \left\| \sum_{k} \xi_k  {\cal A}'_k({\bf x})    \right\| \leq 1\right\}\nonumber\\
=&\mathbb{P}\left\{ \left\| \sum_{k} \xi_k \Big( \frac{1}{\gamma} {\cal A}'_k({\bf x}) \Big)    \right\| \leq \frac{1}{\gamma}\right\}> 1- \epsilon.
\end{align}
This completes the proof. \hfill$\blacksquare$


\subsection{Proof of Theorem \ref{thm:solution3}}

{\bf Proof of Theorem \ref{thm:solution3}:} By setting $\alpha = 2$, it follows from Theorem \ref{thm:conjecture} that with probability at least $1/4$,
\begin{equation}\label{eq:solution3.1}
\|  \widehat{{\bf X}}  \| \leq  \left[\frac{3[\rho_3(\sigma^2+V)]^{\tau_2-1}\cdot \ln(M+N) }{\beta_0\cdot R^{\tau_2-2}}     \right]^{\frac{1}{\tau_\alpha}}.
\end{equation}
For any $1\leq i\leq I$, we have
\begin{equation*}
\widehat{{\bf X}} \bullet \mathcal{B}_i\widehat{{\bf X}} = {\bf B}_i \bullet \widehat{{\bf Y}}^{1/2}{\bf U}^T \bm{\xi} \bm{\xi}^T  {\bf U}  \widehat{{\bf Y}}^{1/2} = \bm{\xi}^T {\bf B}'_i  \bm{\xi},
\end{equation*}
where ${\bf B}'_i  ={\bf U} \widehat{{\bf Y}}^{1/2} {\bf B}_i  \widehat{{\bf Y}}^{1/2} {\bf U}^T  \succeq {\bf 0}$ because ${\bf B}_i  \succeq {\bf 0}$. Then, we can equivalently rewrite 
\begin{equation*}
\widehat{{\bf X}} \bullet \mathcal{B}_i\widehat{{\bf X}}  = \|({\bf B}'_i)^{1/2}\bm{\xi}  \|^2 
= \left\| \sum_{j=1}^{MN}  \xi_j {\rm col}_j [({\bf B}'_i)^{1/2}]  \right\|^2.
\end{equation*}
According to Theorem \ref{thm:conjecture}, for any $i=1,2,\cdots,I$, we have with probability at least $1/4$,
\begin{equation}\label{eq:solution3.2}
\widehat{{\bf X}} \bullet \mathcal{B}_i\widehat{{\bf X}} \leq  \left[\frac{3[\rho_4^{(i)}(\sigma^2+V)]^{\tau_2-1}\cdot \ln(M+N) }{\beta_0\cdot R^{\tau_2-2}}     \right]^{\frac{2}{\tau_2}}.
\end{equation}
Therefore, both of the inequalities \eqref{eq:solution3.1} and \eqref{eq:solution3.2} are valid with probability at least $1-(1/4+1/4) = 1/2$. This completes the proof. \hfill$\blacksquare$


\subsection{Proof of Proposition \ref{prop:aw}}

To prove Proposition \ref{prop:aw}, we first need a preliminary result on the decoupling principle. Although its proof parallels that of \cite[Proposition 1.9]{bourgain1987invertibility},  a proof is given for the sake of completeness.

\begin{lemma}\label{lem:decoupling}
Given an index set $\Omega$ with $|\Omega| <\infty$, let $\{ \xi_k\}_{k\in\Omega}$ be a sequence of centred i.i.d. random variables over a probability $(W, \mathfrak{W},\mu)$ and $(W^{(1)}, \mathfrak{W}^{(1)},\mu^{(1)})$ be an independent copy of $(W, \mathfrak{W},\mu)$. 
For any $\theta \in \mathbb{R}$, then it holds that  
\begin{multline}\label{eq:decoupling}
\mathbb{E} \,{\rm tr} \exp \left( \sum_{j,k\in\Omega} \theta \xi_j \xi_k {\bf A}_j^* {\bf A}_k  \right)\\ \leq \mathbb{E}\, {\rm tr} \exp \left( \mathop{\sum_{j,k\in\Omega}}_{j\not= k} 4\theta \xi_j \xi^{(1)}_k {\bf A}_j^* {\bf A}_k  \right),
\end{multline} 
where $(\xi^{(1)}_1,\cdots,\xi^{(1)}_K)$ over $(W^{(1)}, \mathfrak{W}^{(1)},\mu^{(1)})$ is an independent copy of $(\xi_1,\cdots,\xi_K)$.
\end{lemma}

\begin{IEEEproof}
Let $\{ \eta_k  \}_{k=1}^K$ be a sequence of independent random variable of mean $1/2$ over a probability space $(U,\mathfrak{U},\pi)$ taking only the value $0$ and $1$. Then for any $1\leq j\not = k \leq K$, we have $\int_U \eta_j(u)(1-\eta_k(u)) d\pi(u) = 1/4$. Hence,
\begin{align*}
I =&  \mathbb{E} \,{\rm tr} \exp \left(  \sum_{j,k\in\Omega} \theta \xi_j \xi_k {\bf A}_j^* {\bf A}_k  \right)\nonumber\\
=&\int_W {\rm tr} \exp \Big( \sum_{j,k\in\Omega}\Big[  \int_U \eta_j(u)(1-\eta_k(u)) d\pi(u) \Big] \nonumber \\ 
&\qquad\qquad\qquad\qquad\times4\theta \xi_j(\omega) \xi_k(\omega) {\bf A}_j^* {\bf A}_k  \Big) d \mu(\omega)\nonumber\\
\leq & \int_U \int_W {\rm tr} \exp \Big(  \sum_{j,k\in\Omega}\big[   \eta_j(u)(1-\eta_k(u))  \big]\nonumber\\
&\qquad\qquad\qquad\qquad\times4 \theta \xi_j(\omega) \xi_k(\omega) {\bf A}_j^* {\bf A}_k  \Big)d \mu(\omega) d\pi(u).
\end{align*}
For each $u\in U$, by setting $\sigma(u) = \{j\in \Omega:\; \eta_j(u)=1\}$,
we have 
\begin{align*}
I \leq &  \int_U \int_W {\rm tr} \exp \Big(  \sum_{j\in\sigma(u)} \sum_{k\not\in\sigma(u)}  4\theta \xi_j(\omega) \xi_k(\omega)\nonumber\\
&\qquad\qquad\qquad\qquad\times {\bf A}_j^* {\bf A}_k  \Big) d\pi(u) d \mu(\omega),
\end{align*}
which implies that for each fixed $u\in U$, $\{\xi_j \}_{j\in \sigma(u)}$ are independent of $\{ \xi_k \}_{k\not\in \sigma(u)}$. Hence, we have
\begin{align*}
I \leq & \int_U  \int_W \int_{W^{(1)}} {\rm tr} \exp \Big(  \sum_{j\in\sigma(u)} \sum_{k\not\in\sigma(u)} 4 \theta \xi_j(\omega) \xi_k(\omega^{(1)}) \nonumber\\
&\qquad\qquad\qquad\qquad\times{\bf A}_j^* {\bf A}_k  \Big) d \mu^{(1)}(\omega^{(1)})d \mu(\omega)d\pi(u).
\end{align*}
It suggests that there should exist a $u_0\in U$ such that  
\begin{align*}
I \leq & \int_W \int_{W^{(1)}} {\rm tr} \exp \Big(  \sum_{j\in\sigma(u_0)} \sum_{k\not\in\sigma(u_0)}  4\theta \xi_j(\omega) \xi_k(\omega^{(1)}) \nonumber\\
&\qquad\qquad\qquad\qquad \qquad\qquad\times{\bf A}_j^* {\bf A}_k  \Big) d \mu^{(1)}(\omega^{(1)})d \mu(\omega)\nonumber\\
 \leq & \int_W \int_{W^{(1)}} {\rm tr} \exp \Big(  \sum_{j\in\sigma(u_0)} \sum_{k\not\in\sigma(u_0)}4  \theta \xi_j(\omega) \xi_k(\omega^{(1)}) \nonumber\\
&\qquad\qquad\qquad \qquad\qquad\times{\bf A}_j^* {\bf A}_k +0  \Big) d \mu^{(1)}(\omega^{(1)})d \mu(\omega)\nonumber\\
= & \int_W \int_{W^{(1)}} {\rm tr} \exp \Big(  \Big[\sum_{j\in\sigma(u_0)} \sum_{k\not\in\sigma(u_0)} 4 \theta \xi_j(\omega) \xi_k(\omega^{(1)}) {\bf A}_j^* {\bf A}_k\Big] \nonumber\\
&+ \Big[ \int_W \int_{W^{(1)}}\sum_{j\not\in\sigma(u_0)} \sum_{k\in\sigma(u_0)} 4\theta \xi_j(\omega) \xi_k(\omega^{(1)}) \nonumber\\
&\qquad\qquad\times {\bf A}_j^*{\bf A}_k d \mu^{(1)}(\omega^{(1)})d \mu(\omega) \Big] \Big) d \mu^{(1)}(\omega^{(1)})d \mu(\omega)\nonumber\\
\leq & \int_W \int_{W^{(1)}} {\rm tr} \exp \Big( \mathop{\sum_{j,k\in\Omega}}_{j\not= k}  4 \theta \xi_j(\omega) \xi_k(\omega^{(1)}) \nonumber\\
&\qquad\qquad\qquad\qquad \qquad\qquad\times{\bf A}_j^* {\bf A}_k   \Big) d \mu^{(1)}(\omega^{(1)})d \mu(\omega)\nonumber\\
= & \mathbb{E}\, {\rm tr} \exp \Big( \mathop{\sum_{j,k\in\Omega}}_{j\not= k} 4\theta \xi_j \xi^{(1)}_k {\bf A}_j^* {\bf A}_k \Big),
\end{align*}
where the first equality is derived from the fact that $\{\xi_k\}_{k=1}^K$ are of mean $0$. This completes the proof.
\end{IEEEproof}

{\bf Proof of Proposition \ref{prop:aw}:} 
We first consider the case of even $K$.  Let ${\bf H}_{jk} = ({\bf A}^*_j{\bf A}_k + {\bf A}^*_k{\bf A}_j)$ for any $1\leq j<k\leq K$. Note that $\{{\bf H}_{jk}\}$ are Hermitian. Then, we have 
\begin{align}\label{eq:sumH}
&\sum_{j,k=1}^K \theta \xi_j \xi_k {\bf A}_j^* {\bf A}_k\nonumber\\
 &= \sum_{j} \theta \xi_j\xi_k {\bf A}^*_j{\bf A}_j + \sum_{1\leq j<k\leq K} \theta \xi_j\xi_k {\bf H}_{jk}.
\end{align}
For any even $K\in\mathbb{N}$, we can divide the ${K \choose 2}=K(K-1)/2$ summands $\{\xi_j\xi_k  {\bf H}_{jk} \}_{1\leq j<k\leq K}$ into $K-1$ groups $\{\mathcal{G}_i \}_{i=1}^{K-1}$ such that 1) there are $K/2$ elements in each group; 2) all elements are of the form $(j,k)$ with $j<k$ and 3) the elements in each group can form the index set $\{1,2,\cdots,K\}$.

When $K=6$, the summands $\xi_j\xi_k{\bf A}_j^* {\bf A}_k$ ($1\leq j,k\leq 6,\;j\not=k$) can be divided into $5$ groups: 
\begin{align*}
& \Big\{ \xi_1\xi_2({\bf A}_1^* {\bf A}_2+{\bf A}_2^* {\bf A}_1),\;\xi_3\xi_4({\bf A}_3^* {\bf A}_4+{\bf A}_4^* {\bf A}_3),\;\\
&\qquad\qquad\qquad\qquad\qquad\qquad\qquad\quad\xi_5\xi_6({\bf A}_5^* {\bf A}_6+{\bf A}_6^* {\bf A}_5)   \Big\} \nonumber\\
& \Big\{ \xi_1\xi_3({\bf A}_1^* {\bf A}_3+{\bf A}_3^* {\bf A}_1),\;\xi_2\xi_5({\bf A}_2^* {\bf A}_5+{\bf A}_5^* {\bf A}_2),\;\\
&\qquad\qquad\qquad\qquad\qquad\qquad\qquad\quad\xi_4\xi_6({\bf A}_4^* {\bf A}_6+{\bf A}_6^* {\bf A}_4)   \Big\} \nonumber\\
& \Big\{ \xi_1\xi_4({\bf A}_1^* {\bf A}_4+{\bf A}_4^* {\bf A}_1),\;\xi_2\xi_6({\bf A}_2^* {\bf A}_6+{\bf A}_6^* {\bf A}_2),\;\\
&\qquad\qquad\qquad\qquad\qquad\qquad\qquad\quad\xi_3\xi_5({\bf A}_3^* {\bf A}_5+{\bf A}_5^* {\bf A}_3)  \Big \} \nonumber\\
& \Big\{ \xi_1\xi_5({\bf A}_1^* {\bf A}_5+{\bf A}_5^* {\bf A}_1),\;\xi_2\xi_4({\bf A}_2^* {\bf A}_4+{\bf A}_4^* {\bf A}_2),\;\\
&\qquad\qquad\qquad\qquad\qquad\qquad\qquad\quad\xi_3\xi_6({\bf A}_3^* {\bf A}_6+{\bf A}_6^* {\bf A}_3)   \Big\} \nonumber\\
&\Big \{ \xi_1\xi_6({\bf A}_1^* {\bf A}_6+{\bf A}_6^* {\bf A}_1),\;\xi_2\xi_3({\bf A}_2^* {\bf A}_3+{\bf A}_3^* {\bf A}_2),\;\\
&\qquad\qquad\qquad\qquad\qquad\qquad\qquad\quad\xi_4\xi_5({\bf A}_4^* {\bf A}_5+{\bf A}_5^* {\bf A}_4 )  \Big\}. 
\end{align*}
Since each group can be signified by using the index pairs, we briefly denote as 
\begin{align*}
\mathcal{G}_1 =& \{ (1,2),\, (3,4),\; (5,6)  \} \nonumber\\
\mathcal{G}_2 =& \{(1,3),\, (2,5),\; (4,6)  \} \nonumber\\
\mathcal{G}_3=& \{(1,4),\, (2,6),\; (3,5)  \}  \nonumber\\
\mathcal{G}_4 =& \{(1,5),\, (2,4),\; (3,6)  \}\nonumber\\
\mathcal{G}_5=& \{(1,6),\, (2,3),\; (4,5)  \}. 
\end{align*}
As shown above, the elements in each group can form the entire index set $\{1,2,\cdots,6\}$.

According to \eqref{eq:sumH} and Lemma \ref{lem:decoupling}, we have
\begin{align*}
&\mathbb{E} \,{\rm tr} \exp \left( \sum_{j,k=1}^K \theta \xi_j \xi_k {\bf A}_j^* {\bf A}_k  \right)\nonumber\\
\leq & \mathbb{E} \,{\rm tr} \exp \left( \sum_{1\leq j<k\leq K} 4\theta \xi_j \xi^{(1)}_k {\bf H}_{jk}  \right)\nonumber\\
= & \mathbb{E} \,{\rm tr} \exp \left(\sum_{i=1}^{K-1} \sum_{(j,k) \in \mathcal{G}_i} 4\theta \xi_j \xi^{(1)}_k {\bf H}_{jk}  \right)\nonumber\\
= &\mathbb{E} \,{\rm tr} \exp \left( \sum_{(j,k) \in \mathcal{G}_1 } 4\theta \xi_j \xi^{(1)}_k {\bf H}_{jk}  +\sum_{i=2}^{K-1} \sum_{(j,k) \in \mathcal{G}_i } 4\theta \xi_j \xi^{(1)}_k {\bf H}_{jk}  \right).
\end{align*}

Then, {taking the decoupling principle twice} w.r.t. the index set $\{ \mathcal{G}_i \}_{i=2}^{K-1} $ leads to
\begin{align*}
&\mathbb{E} \,{\rm tr} \exp \left( \sum_{j,k=1}^K \theta \xi_j \xi_k {\bf A}_j^* {\bf A}_k  \right) \leq \nonumber\\
 &\mathbb{E} \,{\rm tr} \exp \left( \sum_{(j,k) \in \mathcal{G}_1 } 4\theta \xi_j \xi^{(1)}_k {\bf H}_{jk}  +\sum_{i=2}^{K-1} \sum_{(j,k) \in \mathcal{G}_i } 4^3\theta \xi^{(2)}_j \xi^{(3)}_k {\bf H}_{jk}  \right)
%
\end{align*}
where the superscripts $^{(2)}$ and $^{(3)}$ stand for the independent copies appearing in the second and the third uses of decoupling principle, respectively. In this manner, the $K/2$ random variables $\{\xi_j\xi^{(1)}_k\}_{(i,j)\in \mathcal{G}_1}$ are independent of each other but also are independent of the others $\{\xi^{(2)}_j\xi^{(3)}_k\}_{(i,j)\in \bigcup_{i=2}^{K-1}\mathcal{G}_i }$. By repeatedly using the decoupling principle, we can obtain independent copies $ \{\zeta_{jk}\}_{j< k}$ of the original dependent random variables $\{\xi_j\xi_k  \}_{j<k}$ as follows: 

\begin{equation*}
\{\zeta_{jk}\}_{(j,k)\in \mathcal{G}_i} =  \big\{  \xi^{(2i-2)}_j \xi^{(2i-1)}_k\big\}_{(j,k)\in \mathcal{G}_i  },
\end{equation*}
for $i=1,2,\cdots,K-1$.
Let ${\bf B}_{jk} = 4^{2K-3} {\bf H}_{jk}$ for any $1\leq j<k\leq K$. Finally, we arrive at
\begin{align}\label{eq:aw.pr1}
&\mathbb{E} \,{\rm tr} \exp \left( \sum_{j,k=1}^K \theta \xi_j \xi_k {\bf A}_j^* {\bf A}_k  \right)\nonumber\\
\leq &\mathbb{E} \,{\rm tr} \left[  \exp \left( \sum_{i=1}^{K-1}\sum_{(j,k) \in \mathcal{G}_i \cup \mathcal{G}'_i} 4^{ 2i-1 }\theta \xi^{(2i-2)}_j \xi^{(2i-1)}_k {\bf H}_{jk} \right)\right]\nonumber\\
= &\mathbb{E} \,{\rm tr} \left[ \sum_{1\leq j< k\leq K} \exp \left( \theta \zeta_{jk} {\bf B}_{jk}  \right)\right]\nonumber\\
\leq & S\cdot \exp\left(  \lambda_{\max} \left( \sum_{1\leq j<k\leq K}\log \mathbb{E} \,{\rm e}^{\theta \zeta_{jk} {\bf B}_{jk}}\right)  \right),
\end{align}
where the last inequality is derived from \cite[Theorem 3.6]{tropp2012user}. Until now, we have completed the proof when even $K$. 

When $K$ is odd, we can introduce an auxiliary term $\xi_{K+1} {\bf A}_{K+1}$ to let the random i.d. series $\sum_{k=1}^K \xi_{k} {\bf A}_{k}$ have even summands, and then the corresponding proof coincides with that of the even $K$ setting. This completes the proof. \hfill$\blacksquare$


\subsection{Proof of Lemma \ref{lem:cs}}

\begin{IEEEproof}
First, we consider the proof of the second inequality in \eqref{eq:cs.lem1}. By Markov's inequality, Lemma \ref{lem:id.mgf} and Proposition \ref{prop:aw}, for any $\theta\in(0,M)$, we have 
\begin{align}\label{eq:cs.pr1}
&\mathbb{P}\left\{ \|{\bf A} {\bf x}\|_2^2 > (1+\delta) \|{\bf x}\|_2^2 ,\;\; \forall \, {\bf x}\in\mathbb{R}^S  \right\}\nonumber\\
 = & \mathbb{P} \left\{ \sigma^2_{\max}({\bf A}) >   (1+\delta) \right\}\nonumber\\
 = & \mathbb{P} \left\{ \lambda_{\max}({\bf A}^* {\bf A}) >   (1+\delta) \right\}\nonumber\\
 = & \mathbb{P} \left\{ {\rm e}^{\theta\lambda_{\max}({\bf A}^* {\bf A})} >  {\rm e}^{\theta (1+\delta)} \right\}\nonumber\\
 \leq &{\rm e}^{-\theta (1+\delta)}\cdot \mathbb{E} {\rm e}^{\theta\lambda_{\max}({\bf A}^* {\bf A})} \nonumber\\
 \leq &{\rm e}^{-\theta (1+\delta)}\cdot \mathbb{E} {\rm tr} \exp \left( \sum_{j,k=1}^K \theta \xi_j \xi_k {\bf A}_j^* {\bf A}_k  \right) \nonumber\\
 \leq &{\rm e}^{-\theta (1+\delta)}\cdot S\cdot \exp\left(  \lambda_{\max} \left( \sum_{1\leq j<k\leq K}\log \mathbb{E} \,{\rm e}^{\theta \zeta_{jk} {\bf B}_{jk}}\right)  \right)\nonumber\\
 \leq&S\cdot \exp\left( - \frac{\bar{\rho}(\bar{\sigma}^2 +\overline{V})}{\overline{R}^2} \cdot Q \left(\frac{\overline{R}(1+\delta)}{\bar{\rho}(\overline{\sigma}^2+\overline{V})}\right) \right)\nonumber\\
 \leq&S\cdot \exp\left( - \frac{\bar{\rho}(\bar{\sigma}^2 +\overline{V})}{\overline{R}^2} \cdot Q \left(\frac{\overline{R}(1-\delta)}{\bar{\rho}(\overline{\sigma}^2+\overline{V})}\right) \right),
\end{align}
where $M$ is defined in Lemma \ref{lem:id.mgf} and the second last inequality is derived by the similar way of proving Theorem \ref{thm:tail} and Corollary \ref{cor:tail1}.

In the similar way, for any $\theta\in(0,M)$, we also arrive at
\begin{align}\label{eq:cs.pr2}
&\mathbb{P}\left\{ \|{\bf A} {\bf x}\|_2^2 < (1-\delta) \|{\bf x}\|_2^2 ,\;\;  {\bf x}\in\mathbb{R}^S  \right\}\nonumber\\
 =& \mathbb{P} \left\{ \sigma^2_{\min}({\bf A}) <   (1-\delta) \right\}\nonumber\\
 = & \mathbb{P} \left\{ \lambda_{\max}(-{\bf A}^* {\bf A}) >  - (1-\delta) \right\}\nonumber\\
 = & \mathbb{P} \left\{ {\rm e}^{\theta\lambda_{\max}(-{\bf A}^* {\bf A})} >  {\rm e}^{-\theta (1-\delta)} \right\}\nonumber\\
 \leq &{\rm e}^{\theta (1-\delta)}\cdot \mathbb{E} {\rm e}^{\theta\lambda_{\max}(-{\bf A}^* {\bf A})} \nonumber\\
 \leq &{\rm e}^{\theta (1-\delta)}\cdot \mathbb{E} {\rm tr} \exp \left( \sum_{j,k=1}^K -\theta \xi_j \xi_k {\bf A}_j^* {\bf A}_k  \right) \nonumber\\
 \leq &{\rm e}^{\theta (1-\delta)}\cdot S\cdot \exp\left(  \lambda_{\max} \left( \sum_{1\leq j<k\leq K}\log \mathbb{E} \,{\rm e}^{\theta \zeta_{jk} (-{\bf B}_{jk})}\right)  \right)\nonumber\\
 \leq&S\cdot \exp\left( - \frac{\bar{\rho}(\bar{\sigma}^2 +\overline{V})}{\overline{R}^2} \cdot Q \left(\frac{\overline{R}(1-\delta)}{\bar{\rho}(\overline{\sigma}^2+\overline{V})}\right) \right),
\end{align}
where the last inequality is obtained by the same way to achieve Theorem \ref{thm:tail} and Corollary \ref{cor:tail1}. The combination of \eqref{eq:cs.pr1} and \eqref{eq:cs.pr2} leads to the final result \eqref{eq:cs.lem1}. This completes the proof. 
\end{IEEEproof}


\subsection{Proof of Theorem \ref{thm:rip}}

\begin{IEEEproof}
As shown in Lemma \ref{lem:cs}, for each $\mathcal{I} \subset \{1,\cdots,D\}$ with $|\mathcal{I}| =S$, the $M\times S$ random i.d. series $[{\bf A}]_\mathcal{I}$ fails to satisfy the RIP \eqref{eq:cs.lem1} with probability at most
\begin{equation}\label{eq:rip.pr1}
2S\exp\left( - \frac{\bar{\rho}(\bar{\sigma}^2 +\overline{V})}{\overline{R}^2} \cdot Q \left(\frac{\overline{R}(1-\delta)}{\bar{\rho}(\overline{\sigma}^2+\overline{V})}\right) \right).
\end{equation}
Since there are ${D \choose S}\leq ({\rm e} D/S)^S$ possibilities to select $\mathcal{I}$ from $\{1,\cdots,D\}$ and the expression \eqref{eq:rip.pr1} is a decreasing function w.r.t. $\bar{\rho}>0$, the ${\rm RIP_S}(\delta)$ \eqref{eq:rip1} will fail to hold with probability at most 
\begin{equation}\label{eq:rip.pr2}
2S\cdot ({\rm e} D/S)^S\cdot \exp\left( - \frac{\bar{\rho}_S(\bar{\sigma}^2 +\overline{V})}{\overline{R}^2} \cdot Q \left(\frac{\overline{R}(1-\delta)}{\bar{\rho}_S(\overline{\sigma}^2+\overline{V})}\right) \right).
\end{equation}
Therefore, if the constants $c_1,c_2>0$ satisfy Conditions \eqref{eq:rip.cond1} and \eqref{eq:rip.cond2}, then the expression \eqref{eq:rip.pr2} will smaller than $(\log 2S)-c_2M$. This completes the proof. 
\end{IEEEproof}

\section*{Acknowledgment}
The authors would like to thank the anonymous reviewers and the editors for their valuable comments and suggestions. 

\bibliographystyle{IEEEtran}

\bibliography{ref-iddrm}

\end{document}